\documentclass{aastex63}
\usepackage{graphicx}
\usepackage{amssymb}                    % useful mathematical symbols
\usepackage{ulem}
\usepackage{float}
\usepackage{rotating}
\usepackage{enumitem}
\usepackage{rotating}
\usepackage{makecell}
\usepackage{appendix}
\usepackage{fixmath}
\usepackage{color}
\usepackage{ulem}
\usepackage{threeparttable}

\newcommand{\x}{\ensuremath{\times}}

\usepackage{times} 
\usepackage{amsmath}
\submitjournal{ApJ}

\shortauthors{Athiray et al.}
\shorttitle{{\magixs} Flight Calibration II}

\newcommand{\xrt}{\textit{Hinode}/XRT}

\newcommand{\magixs}{\textit{MaGIXS}}

\newcommand{\degree}{$^{\circ}$}

\newcommand{\usra}{NASA Postdoctoral Program, NASA Marshall Space Flight Center, ST13, Huntsville, AL, USA}
\newcommand{\msfc}{NASA Marshall Space Flight Center, Huntsville, AL 35812, USA}
\newcommand{\izentis}{Izentis LLC, PO Box 397002, Cambridge, MA 02139}
\newcommand{\sao}{Center for Astrophysics $\vert$\ Harvard\ \&\ Smithsonian, 60\ Garden\ Street, Cambridge, M\!A \ 02138}
\newcommand{\lbnl}{Center for X-Ray Optics, Lawrence Berkeley National Laboratory, Berkeley, CA 94720}
\newcommand{\themit}{Space Nanotechnology Laboratory, MIT Kavli Institute for Astrophysics and Space Research, Massachusetts Institute of Technology, Cambridge, MA 02139}
\usepackage[symbol]{footmisc}
\renewcommand{\thefootnote}{\fnsymbol{footnote}}
\begin{document}

%%------------------------------------------------------------------------
%%
%%						TITLE
%%
%%------------------------------------------------------------------------

\title{Calibration of the {\magixs} experiment II: Flight Instrument Calibration}
\correspondingauthor{P.S. Athiray}
\email{athiray.panchap@nasa.gov}
\author[0000-0002-4454-147X]{P.S. Athiray}\affiliation{\usra}

\author[0000-0002-5608-531X]{Amy R.\ Winebarger}\affiliation{\msfc}
\author[0000-0002-7139-6191]{Patrick Champey}\affiliation{\msfc}
\author{Ken Kobayashi}\affiliation{\msfc}
\author[0000-0002-6172-0517]{Sabrina Savage}\affiliation{\msfc}
\author{Brent Beabout}\affiliation{\msfc}
\author{Dyana Beabout}\affiliation{\msfc}
\author{David Broadway}\affiliation{\msfc}
\author[0000-0001-5927-3300]{Alexander R. Bruccoleri}\affiliation{\izentis}
\author{Peter Cheimets}\affiliation{\sao}
\author{Leon Golub}\affiliation{\sao}
\author{Eric Gullikson}\affiliation{\lbnl}
\author{Harlan Haight}\affiliation{\msfc}
\author[0000-0001-9980-5295]{Ralf K.\ Heilmann}\affiliation{\themit}
\author{Edward Hertz}\affiliation{\sao}
\author{William Hogue}\affiliation{\msfc}
\author{Steven Johnson}\affiliation{\msfc}
\author{Jeffrey Kegley}\affiliation{\msfc}
\author{Jeffery Kolodziejczak}\affiliation{\msfc}
\author{Chad Madsen}\affiliation{\sao}
\author[0000-0001-6932-2612]{Mark L.\ Schattenburg}\affiliation{\themit}
\author{Richard Siler}\affiliation{\msfc}
\author[0000-0002-7219-1526]{Genevieve D.\ Vigil}\affiliation{\msfc}
\author{Ernest Wright}\affiliation{\msfc}

\begin{abstract}
The Marshall Grazing Incidence X-ray Spectrometer (\magixs) is a sounding rocket experiment that  observes the soft X-ray spectrum of the Sun from 6.0 - 24 {\AA} (0.5 - 2.0 keV), successfully launched on 30 July 2021. End-to-end alignment of the flight instrument and calibration experiments are carried out using the X-ray and Cryogenic Facility (XRCF) at NASA Marshall Space Flight Center. In this paper, we present the calibration experiments of {\magixs}, which include wavelength calibration, measurement of line spread function, and determination of effective area.  Finally, we use the measured instrument response function to predict the expected count rates for {\magixs} flight observation looking at a typical solar active region. 
\end{abstract}
\keywords{Sun:corona, X-rays}
\section{Introduction} 
\label{sec:intro}
Understanding what drives the physics of coronal heating remains one of the critical problems of solar physics. Though many different theories have been put forth \citep[e.g.,][]{cranmer2019}, two of the most likely competing physical mechanisms are sporadic energy release through magnetic reconnection \citep{Parker1988, cargill1995, Klimchuk2006} and steadier energy release through dissipation of Alfven waves \citep[e.g.,][]{asgaritarghi2012}.  One of the fundamental differences between these mechanisms is the frequency of heating events (1/$\tau_{heat}$), where $\tau_{heat}$  is the average time between two heating events on a loop strand. The evolution of the plasma temperature and density depend strongly on $\tau_{heat}$.  For high frequency heating, where the time between heating events is much shorter than the cooling time of the plasma, the temperature and density remain relatively constant, while for low frequency heating, where the time between heating events is long compared to the cooling time, they evolve dynamically. For cases where loop strands are not resolved, different heating scenarios result in different predicted distributions of the temperature of the plasma.  A proxy for the distribution of the plasma temperature is called the differential emission measure (DEM) distribution, or its integral form, emission measure distribution (EM), which would be an observational discriminator to distinguish the mechanisms. A broad DEM with the presence of higher than average coronal temperatures would imply low-frequency heating, while high-frequency heating would yield a narrow DEM \citep{klimchuk2017}. Measuring the high temperature EM slope is found to be the ``smoking gun'' observation required to constrain the timescale of heating events \citep[e.g.,][]{barnes2016a, barnes2016b,barnes2019, reep2013a}.  Although some  limited diagnostics are currently available to discriminate high-temperature emission \citep[e.g.,][]{Warren_2012}, most current space instrumentation has a so-called ``blind-spot'' to the crucial  low emission measure, high-temperature plasma above 5 MK \citep{winebarger2012}.  \cite{Athiray2019} demonstrated that the ratios of emission lines from Fe XVII, XVIII, and XIX are particularly sensitive to the high temperature EM slope and provide an excellent diagnostic of coronal heating frequencies. 

The Marshall Grazing Incidence X-ray Spectrometer (\magixs) is a sounding rocket experiment developed by the NASA Marshall Space Flight Center (MSFC) and the Smithsonian Astrophysical Observatory (SAO), launched successfully on 30 July 2021 from the White Sands Missile Range, NM. {\magixs} is designed to observe, for the first time, soft X-ray (SXR) emission of high-temperature, low-emission plasma of spatially and spectrally-resolved solar coronal structures  in the energy range from 0.5 - 2.0 keV \citep{kobayashi2010,kobayashi2018, champey2016}. The {\magixs} spectrometer is sensitive to a series of warm and hot plasma emission lines simultaneously through the same optical path, providing spatial information along a slot. The bandpass includes several emission lines formed by the key ionization stages of Fe (XVII, XVIII, and XIX), which will extend the DEM coverage from 3MK to 10MK and thereby help to constrain the slope of the high-temperature EM fall-off \citep[e.g.,][]{Athiray2019}.

Because the determination of the high temperature fall off requires ratios of spectral lines formed at different wavelengths, knowledge of the relative calibration and potential errors in calibration are required.  We were particularly motivated to constrain the radiometric calibration in the 10-17\,\AA\ region, which contains several of these important diagnostic lines. Therefore, we have performed a detailed characterization of {\magixs} using targets that emit X-ray lines in the above wavelength range, at the NASA MSFC X-ray and Cryogenic Facility (XRCF). Prior to calibration tests, we have established an optimized method for photon counting in CCD images, and applied the same to characterize the X-ray source at XRCF, which is used for {\magixs} calibration. These results are published in \citet{Athiray2020}, here onwards referred to as Paper I.
 
In this paper, we present the experiments, analysis, and results of the ground calibration of {\magixs} carried out at the XRCF. Section~\ref{sec:design} provides a description of the {\magixs} instrument design, a brief overview of the optical components, and a short summary on the alignment and testing of the X-ray mirrors. In Section \ref{sec:setup} we describe the calibration experiment test setup and provide a summary of data collection. Methods of data analysis and results including wavelength calibration, point spread function, estimation of throughput flux and determination of effective area of {\magixs} are detailed in Section  \ref{sec:wavecalibration}. In Section \ref{sec:flightpredict} we describe the predicted expected count rates from the {\magixs} flight for a typical active region using the calibration products, and we conclude with an overall summary and discussion in Section \ref{sec:summary}.

\section{{\magixs} Instrument Overview} \label{sec:design}
\subsection{Instrument Description}

\begin{figure}[h]
    \centering
    \includegraphics[width=0.8\linewidth]{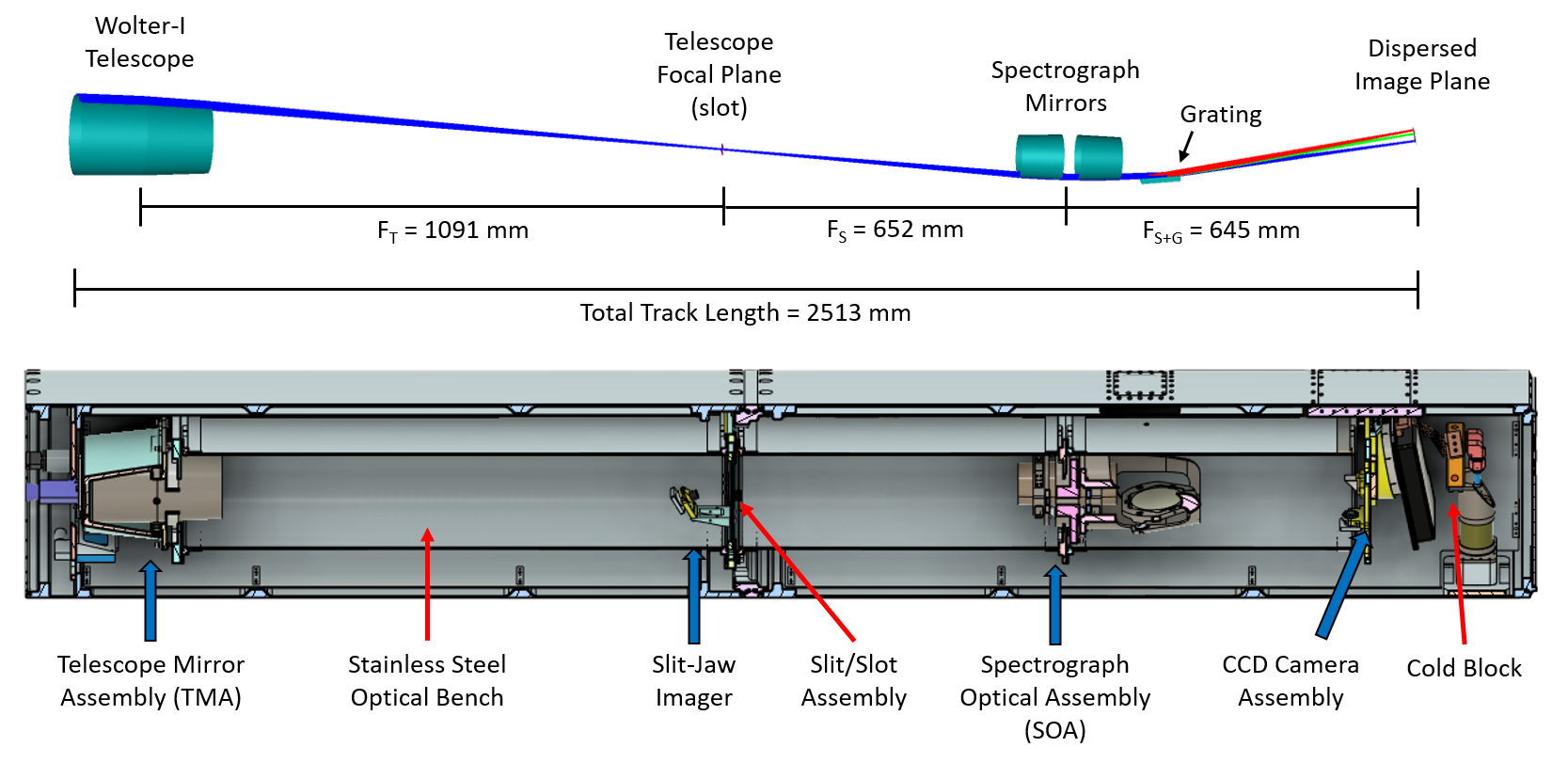}
    \caption{The optical layout and ray trace of the {\magixs} instrument design . All the major optical components, excluding the slit-jaw imaging system, filters, and science camera are listed in the figure.}
     \label{fig:design}
\end{figure}

The {\magixs} optical instrument is described in detail in \cite{kobayashi2014} and \cite{Champey2015} and reviewed briefly here.  The instrument includes a single shell Wolter-I type telescope mirror, a slot with slit-jaw context imager, a pair of conjugate parabolic spectrometer mirrors (SM1 and SM2), a  grating, and a CCD detector. The design also includes several small reference mirrors and alignment reticles, which are used for co-alignment of optical elements. Figure \ref{fig:design} shows the optical layout of {\magixs} with all the key optical components distinctly marked. The top panel in Figure \ref{fig:design} shows the optical path in X-rays. Table \ref{tab:opticalcomponents} summarizes the specifications of the individual optical components of {\magixs}.

\begin{deluxetable}{lclll}
 \tablecaption{Summary of description of {\magixs} optical components. \label{tab:opticalcomponents}}
 \tablehead{
              \colhead{S.No}   & \multicolumn{2}{c}{Optical Component} & \colhead{Description} & \colhead{Remarks}}
              \startdata
                  1 & Telescope Mirror  & Entrance Filter & 50 nm of Al with (assumed) 10 nm of Al$_{2}$O$_{3}$  &   Test Configuration : No filter \\
                               & Assembly (TMA)                &  & mounted on 70 lpi Ni mesh & Flight Configuration : Filter Present \\
                               &                   &  &  &  \\
                               &                   &Telescope Mirror  & Mirror type : Wolter-I, Ni replicated &  Test Configuration : Spacers for source \\
                               &                   &  & Graze angle : 1{\degree} & at finite distance \\
                               &                   &  & Surface roughness : 0.5 nm &  Flight Configuration : Spacers for source \\
                               &                   &  & Focal length : 1.09~m &  at infinity\\
                               &                   &  & Geometric area (A$_{geo}$) : 0.87 cm$^2$ &  \\
                               &                   &  & Coating : Iridium, 40 nm thick &  \\
                  \hline
  2               &  Slot & & Molybdenum film on circular Aluminum mount, & Test Configuration : Custom slot with   \\
                  &   & &3.84~mm wide, 10.5~mm long ,  & large pinholes (see Figure \ref{fig:slot}, text)  \\  
                  &   & &Circular Aluminum mount 25~mm diameter  & Flight Configuration : 12{\arcmin}wide, 33 {\arcmin} long  \\
                  \hline
  3               & Spectrometer Optical & Spectrometer Mirrors  &Mirror type : Paraboloid Ni replicated &    \\
                  &             Assembly (SOA)       &  SM1, SM2 & Graze angle : 2 {\degree} &  \\
                  &                   &  & Surface roughness : 0.5 nm &  \\
                  &                   &  & Focal length : 0.79~m &  \\
                  &                   &  & Coating : Iridium, 40 nm thick &  \\
                  &                   &  &  &                             \\
                  &                   &  &  Graze angle : 2{\degree} & \\
                  &                   & Grating &  Blaze angle : 1.6{\degree} & Efficiency measured at Lawrence  \\
                  &                   &  &  Coating : Iridium, 5~nm thick  &Berkeley National Laboratory\\
                  &                   &  &  73 mm long  &\\
                  &                   &  &  Varied line spacing (d$_{center}$ : 4760 {\AA})  &\\
              \hline
4                 &  Focal Plane Filter & & 150 nm Al with (assumed) 10 nm Al$_{2}$O$_{3}$ &    \\
                  &                     & & mounted on 200 nm polyimide &    \\
                \hline
5                 &  Detector           & & CCD 2k $\times$ 1k pixels, 15 $\mu$m pixel size &  \\
\enddata
\end{deluxetable}

{\magixs} is the first instrument to fly X-ray mirrors that were polished using a state-of-the-art deterministic polishing technique, which was shown to significantly minimize figure errors of the mandrels \citep{Davis2019, champey2019}.  Because only a segment of each of the reflective surfaces (see ray trace Figure \ref{fig:design}) directs rays to the grating, deterministic polishing only occurred over a 43$^{\circ}$ section of each mandrel. The 73~mm long, varied line space (VLS), planar diffraction grating, fabricated by Izentis LLC, diffracts $\approx$ 34 $^{\circ}$ of the full system aperture, dispersing primarily the first order diffracted X-ray beam and second order beam to the science camera. The X-ray mirrors, including telescope and spectrometer (SM1, SM2), and grating are mounted on appropriate structures to form different assembly modules, which was carried out at SAO. The telescope mirror assembly (TMA) carries the Wolter-I optic, and the arrangement of spectrometer mirrors aligned to the grating is collectively called the Spectrometer Optical Assembly (SOA). A thorough description of the assemblies (TMA, SOA) along with the adopted alignment methods are presented in \cite{Ed2020}. 

The science camera has been developed by MSFC for suborbital missions \citep{Laurel2019} and utilizes a back-illuminated, ultra-thinned, astro-processed Te2V CCD.  The camera is positioned off axis to collect chiefly the first-order diffracted spectral image. In addition to the X-ray optical system, {\magixs} also includes a slit-jaw context imager to aid in pointing during flight and alignment to other data sets after flight.  The slit-jaw system provides context solar image during the flight.

Finally, the {\magixs} instrument includes an entrance and focal plane filter.  The entrance filter is 50 nm of Al mounted on mesh, while the focal plane filter is 150 nm of Al mounted on polyimide.  The entrance filter was chosen to allow for EUV light to be focused at the focal plane of the TMA and reimaged by the slit-jaw optics, while the focal plane filter was chosen to reduce the out of band light on the detector.  For purposes of the effective area calculation, we assume the surfaces of the filters have oxidized and include a 10 nm thickness of Al$_{2}$O$_{3}$.

\subsection{Instrument  Configurations}

For the end-to-end alignment and calibration tests, {\magixs} was built sequentially from the telescope end to the science camera end, adding each optical component one-by-one in the `test configuration', which is slightly different from the `flight configuration'. We highlight the differences between the two configurations in the remarks column in Table \ref{tab:opticalcomponents}. After adding each component, the built-up instrument was aligned using a theodolite and reference mirrors and reticles, and confirmed by X-ray measurements at the XRCF, which is a 500~m evacuated X-ray beamline. The assembly tests were designed to confirm the focal position of the TMA and to verify the alignment of the overall integrated instrument. Table \ref{tab:configs} gives a broad overview of the test series performed at XRCF and the corresponding instrument setup. A paper summarizing the alignment strategy employed for MaGIXS 
%along with the results obtained during the flight campaign 
will be presented elsewhere (Champey et al, in preparation). These alignment checks also provide data we can use for component level calibration of the X-ray mirrors, discussed in Section~\ref{sec:compcal}.  Additionally, before integration into the SOA, the grating was tested independently at LBNL.  The results of that test is also given in Section~\ref{sec:compcal}.  

\begin{figure}[h]
    \centering
    \includegraphics[width=0.3\linewidth]{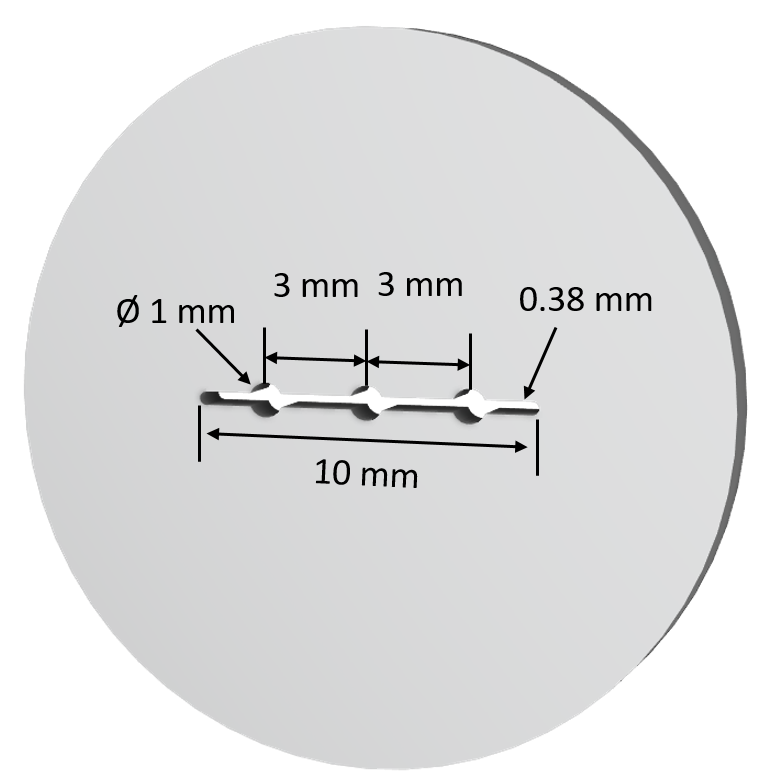}
    \caption{Sketch of the custom slot made with three large pin holes used for the calibration tests.}
     \label{fig:slot}
\end{figure}

{\magixs} was originally designed with a narrow slit at the focus position of the TMA.  Soon after we conducted X-ray alignment tests with TMA, we realized that it was difficult to co-align the instrument with the original slit that was aimed to achieve 6{\arcsec{}} angular resolution. However, due to the degraded mirror point spread function and the advancement of unfolding techniques for slot imaging spectrograph data \citep{winebarger2019}, it was determined to replace the slit with a 12{\arcmin} slot during integration and calibration.  For the test configuration, we fabricated a custom made slot with three large pinholes drilled on an Aluminum slit, with each $\sim$ 1~mm size, one at the center of the slot (on-axis) and two on either sides to the center of the slot (off-axis) as shown in Figure \ref{fig:slot}. The off-axis pinholes are about 9.5{\arcmin}  spaced from the center hole. For the flight configuration, the custom designed pinhole slot will be replaced with a newly fabricated lumogen coated slot mask with $\sim$ 3.84~mm ie. $\sim$ 12{\arcmin} wide and 33{\arcmin} long. We mention that results of {\magixs} calibration in the test configuration can be applied to the flight configuration, as changes are minimal (see Table \ref{tab:opticalcomponents} columns 4,5) and does not affect the calibration parameters. 

\begin{deluxetable}{l l l l }
\tablecaption{Summary of test series and instrument arrangement. \label{tab:configs}}
\tablehead{
\colhead{S.No}&\colhead{Test series} & \colhead{Instrument setup }&\colhead{Remarks}}
\startdata
1& TMA focus check & Source - TMA - Camera & Series of experiments to evaluate On/Off axis TMA performance \\
2& Alignment check & Source - TMA - Slot - SM1 - SM2 -Camera & Series of experiments at out-of-focus positions to verify alignment\\
3&Calibration &Source - TMA - Slot - SOA - Camera & Observed a shift in SOA position during the first round of calibration experiments\\
& & & adjusted SOA position performed final calibration described in this paper
\enddata
\end{deluxetable}

\section{{\magixs} Calibration} \label{sec:setup}

The goal of {\magixs} calibration is to determine the wavelength calibration, assess the line spread and point spread function, and measure the actual effective area of the instrument.  In this section, we describe the calibration experiment set up, the sources used, and the summary of data collected.  

\subsection{Experiment setup} 
\begin{figure}[h]
    \centering
    \includegraphics[width=0.8\linewidth]{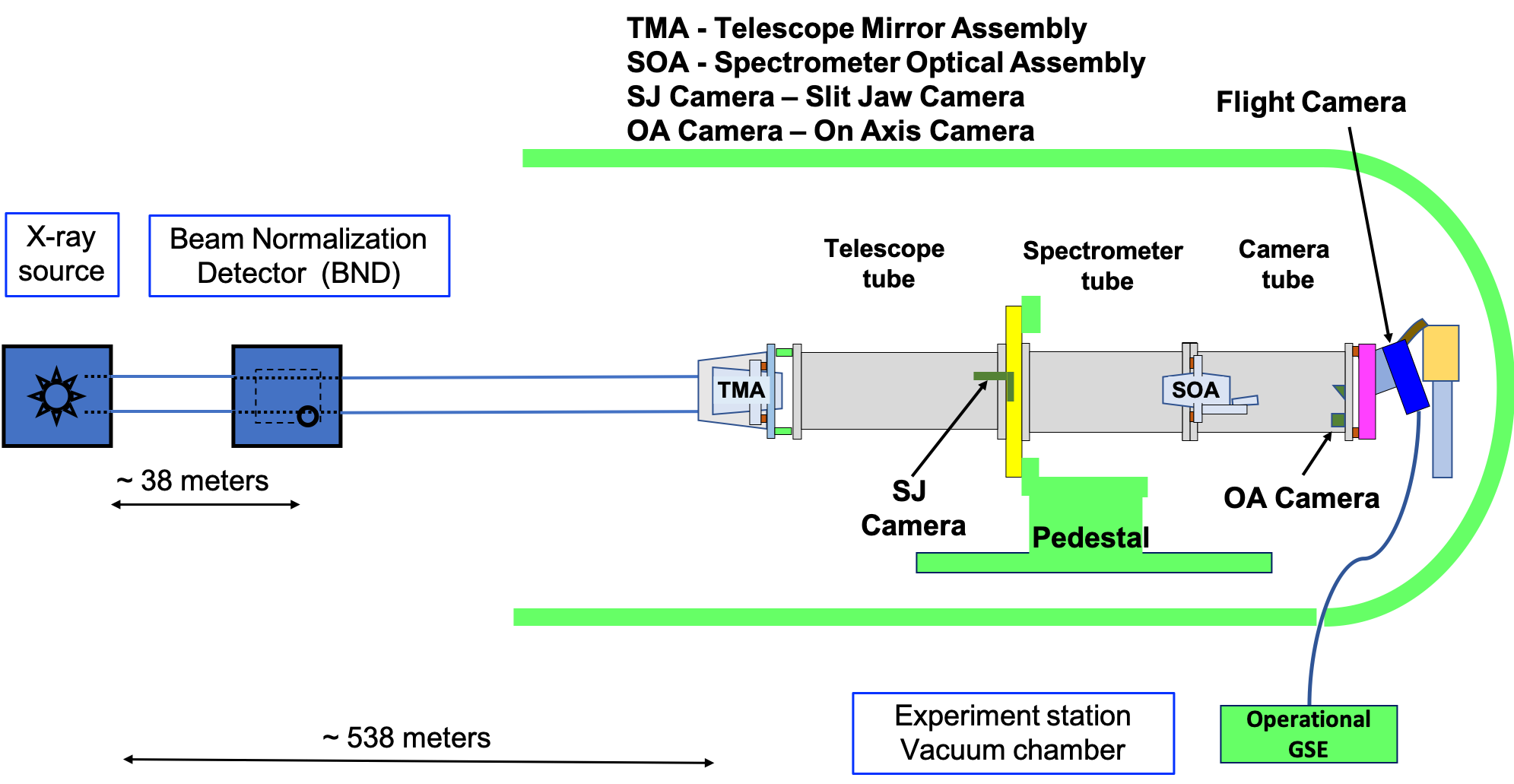}
    \caption{Schematic view of the test setup for {\magixs} calibration tests at the X-ray and Cryogenic Facility (XRCF) at NASA MSFC. In all test configurations the incident source flux is monitored by a beam normalization detector (BND).}
    \label{fig:setup}
\end{figure}

Figure \ref{fig:setup} shows the schematic of {\magixs} calibration set up assembled in the test configuration at the XRCF. The aligned, built up instrument (TMA-slot-SOA-camera) is placed on the five axis mount stage on the experiment station inside the XRCF vacuum chamber. The entrance of TMA is placed at $\approx$ 536~m distance from the X-ray source. The X-rays from the source are transmitted through an evacuated  518~m long guide tube to an evacuated 7.3~m diameter $\times$ 22.9~m long instrument chamber. An optical laser is also placed at the source end next to the X-ray source, which is used to provide rough alignment between the source and the telescope in air.  This laser could be seen on the slot surface at the focus of the TMA using the slit-jaw imager.  The flight camera is placed at the off-axis detector plane to measure the wavelength dispersed image coming from the grating.  The CCD is mounted in a copper carrier that is connected via a copper strap to a cold block. The cold block is actively cooled to roughly -100$^{\circ}$\,C using liquid nitrogen, which results in a CCD carrier average temperature of roughly -70$^{\circ}$\,C.  The carrier temperature is controlled such that it is maintained constant within $\pm$~5$^{\circ}$\,C. During one of our calibration test run the cooling system was paused inadvertently and resulted in an increased noise in the CCD, which is described in section \ref{sec:preprocess}

The facility provided a beam normalization detector (BND) to monitor the incident X-ray spectrum and flux. The BND is a flow proportional counter (FPC) that uses flowing P10 gas (90\% Argon, 10\% Methane) at a pressure of 400 Torr. For a detailed description of the FPC, see \citet{Bradford1997}. The BND is mounted on a translation stage within the guide tube at a distance of 38 meters from the source, and outside of the Electron Impact Point Source (EIPS)-CCD beam path.  Apertures of appropriate diameters are used in front of the BND to avoid saturation. In Paper I, we established that the flux measured from BND is reliable to within 20\% and is consistent with the values reported in literature. We employed the same BND to simultaneously monitor the incident flux during the end-to-end alignment tests as well as for the calibration tests. 
{}

\subsection{X-ray Source}
 For calibration tests, the X-ray source at XRCF was operated with an anode voltage that is 4~$\times$ the L-shell or 5~$\times$ the K-shell binding potential of the target at a permissible current. A list of the targets, used for the end-to-end alignment tests and calibration tests is given in Table~\ref{tab:calib_targets}. No filter was placed in front of the X-ray source for the calibration tests. The unfiltered X-ray source spectrum comprises target lines and bremsstrahlung continuum. The X-ray source operating parameters such as anode voltage and current (Columns 4 and 5) are chosen such that the expected photon flux on CCD is relatively low to perform `photon counting' and also to avoid saturation and pile up artifacts in the resulting image. 
 
 \begin{deluxetable}{c c c c c c c c}
\tablecaption{List of calibration targets and respective line energies. \label{tab:calib_targets}}
\tablehead{
\colhead{S.No}&\colhead{Target} &\colhead{Line energy}& \colhead{Voltage}& \colhead{Current} & \colhead{BND aperture diameter} & \colhead{Test} &\colhead{Source} \\
\colhead{}&\colhead{}  &\colhead{keV}& \colhead{kV}& \colhead{mA} &\colhead{cm} &&\colhead{Filter}}
\startdata
1&Ni-L& 0.85& 3.40 & 5.8  &0.1 & Calibration & No filter\\
2&Zn-L & 1.01& 4.00 & 3.0  &0.1 & Calibration & No filter\\
3&Mg-K \footnote[1]{This data set contained additional noise} & 1.25& 6.50 & 1.5 &0.1 & Calibration& No filter\\
4&Mg-K & 1.25& 2.60 & 3.0 &0.4 & End-to-end - TMA focus check  & K2\\
5&Mg-K & 1.25& 2.60 & 3.0 &0.4 & End-to-end - Alignment check & K2
\enddata
\end{deluxetable}

\subsection{Data Summary and Pre-processing}
\label{sec:preprocess}

\begin{figure}[h]
    \centering
    \includegraphics[width=0.99\textwidth]{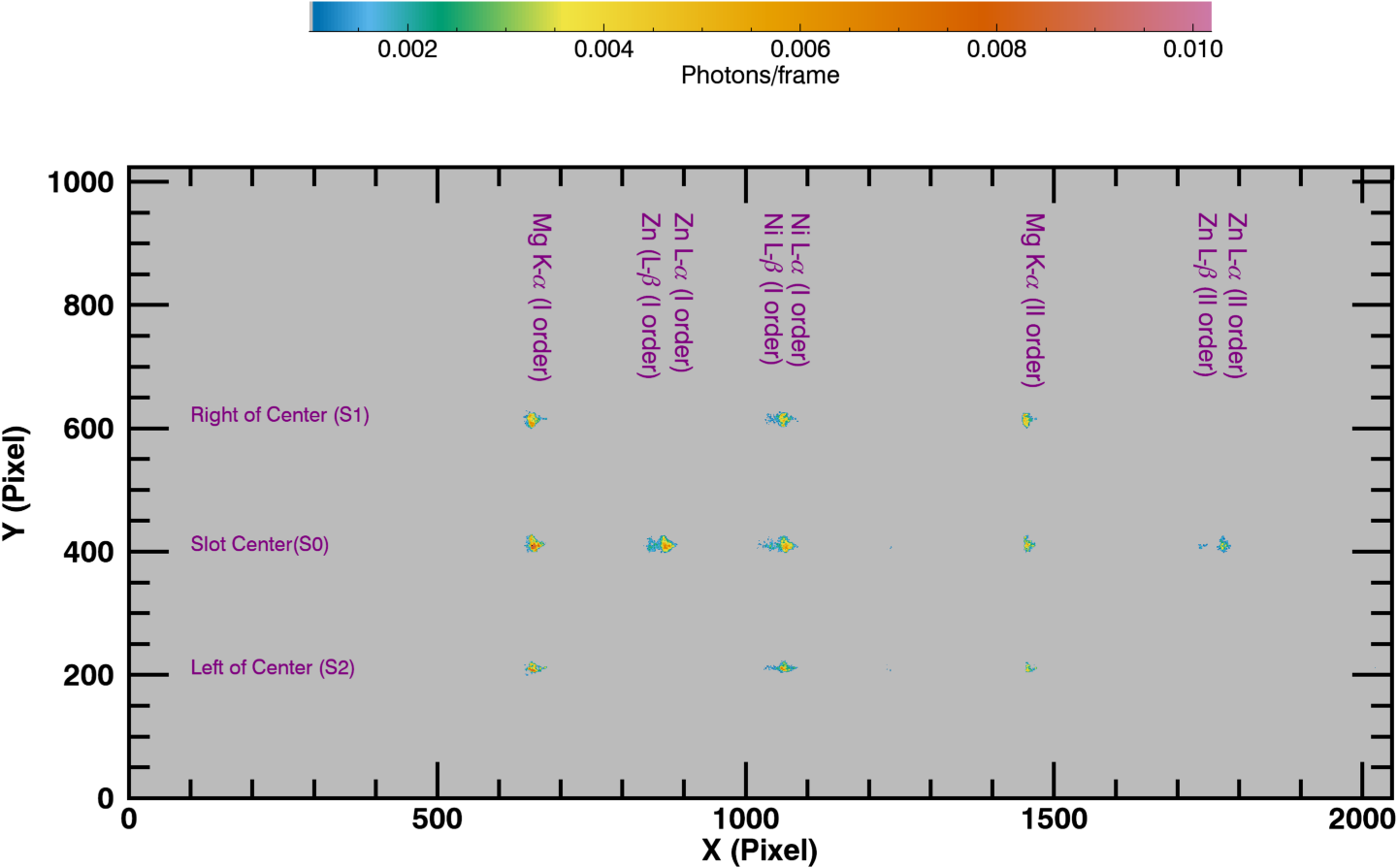}
    \caption{{\magixs} X-ray image summarizing all data collected from the calibration tests normalized with number of data frames at each target and slot position. This image is smoothed over three pixels for better data visualization. The data collection includes measurement at three beam positions S0, S1, and S2 using three targets Mg, Zn, and Ni respectively, which are labeled.  The second orders are dispersed toward the right side of the figure.  The data have been event processed, which includes determining photon energy deposits and corresponding photon hit locations to correctly include multi-pixel photon hits. 
    }
    \label{fig:uncalibrated image and spectra}
\end{figure}

The converged X-ray beam from TMA was allowed to pass through the pinhole slot at three different positions along the slot viz on-axis S0 (center), and off-axis S1, S2 (either sides of the slot center) respectively. Calibration data were acquired for different targets using the science camera operated in the frame transfer mode with an exposure time of $\sim$3~sec per frame.  More than 600 frames of data were acquired at each beam position for adequate statistics. Data for Zn target was collected only at S0 position due to limited beam time availability. Dark frames were acquired at regular intervals by closing a gate valve between the source and the {\magixs} telescope. Standard CCD data reduction routines are followed to remove bias level, dark current, and fixed pattern noise in the images. The measured gain of the camera was determined prior to calibration tests using a sealed radioactive source Fe$^{55}$ and found to be $\sim$ 2.66 electrons per data number (DN). The measured RMS read noise is approximately 10 e$^{-}$, which includes read and dark noise. The camera cooling was inadvertently halted during the measurement of data set \#3 in Table \ref{tab:calib_targets}. The subsequent rise in temperature introduced additional dark noise and hot pixels to that data set. Extra care was taken while processing the data to isolate hot pixels. We note the data set is still useful for wavelength calibration, however the additional noise complicates the ability to precisely determine the photon energy and hence has limited usefulness for radiometric calibration. Figure \ref{fig:uncalibrated image and spectra} shows the summary of collated, uncalibrated, pre-processed {\magixs} data collected from the calibration tests, with the prominent target emission lines and beam positions labelled. This image is a combination of data collected at each target and beam position on the slot, smoothed over three pixels for better visualization.

{}

\section{Data Analysis}
\label{sec:wavecalibration}
The calibration data are analyzed using the optimized event selection algorithm described in Paper I, which identifies individual photon hit locations and calculates the total energy deposited by each photon on the detector. In the case of multipixel events, the algorithm sums adjacent shared pixels to reconstruct the photon energy, and the pixel where the maximum energy is deposited is considered for image construction.  Figure \ref{fig:uncalibrated image and spectra} was constructed using this algorithm.  The {\magixs} is a spectrograph with a diffraction grating, which means photons with different energies (wavelengths) are dispersed and then detected at different pixel coordinates along the dispersion axis on the CCD. Hence, we can measure photon energy in two ways: (i) energy deposited on the detector from a photon event recovered from the event selection algorithm and (ii) the location where photon falls on the detector in the dispersion direction `X', which can be determined through pixel-to-wavelength calibration.

\subsection{Calculating Slot-SOA-Camera Roll}

\begin{figure}
    \centering
    \includegraphics[width=0.8\textwidth]{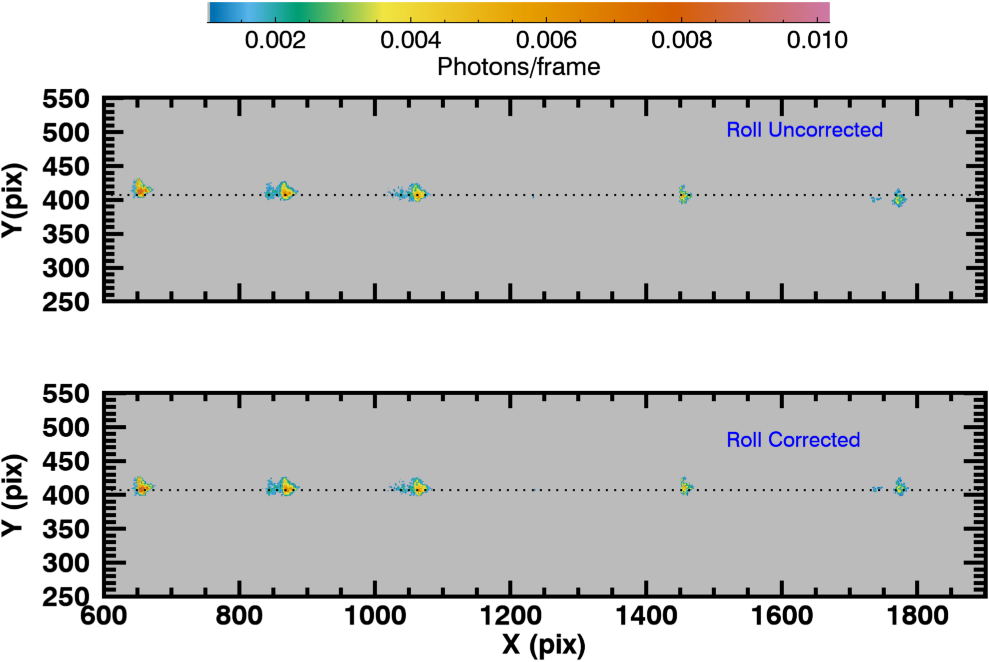}
    \caption{An example plot comparing the cropped S0 image for all targets with and without roll correction applied to the data.}
    \label{fig:roll}
\end{figure}

Using the event processed X-ray images, for each target/pointing data combination, we created summed spectra along the dispersion axis (X) and along the cross-dispersion  axis (Y). These summed spectra are fitted with Gaussian functions to determine the mean (X,Y) pixel locations where the peak of the distribution occurs. We find the peaks in X  are well aligned at each target line, for different pointing S0, S1 and S2. This alignment implies that we do not find any roll in the slot orientation with respect to the SOA and camera. 
However, the locations of the peaks in Y showed a small deviation from parallel to the detector pixel rows at each pointing. This misalignment indicates that there is a small roll angle involved between the SOA and camera, which we determined to be ~-0.64{\degree} from the detector X  axis. For example, Figure \ref{fig:roll} shows the uncorrected S0 image plotted on the top panel and the roll angle corrected S0 image plotted in the bottom panel. With this correction, we note how well the spots align with the dispersion (X) axis of the detector indicated by the solid horizontal line.  All event processed data are first corrected for this roll angle before the analysis for calibration.

\subsection{Wavelength calibration}
\begin{figure}
    \includegraphics[width=0.5\textwidth]{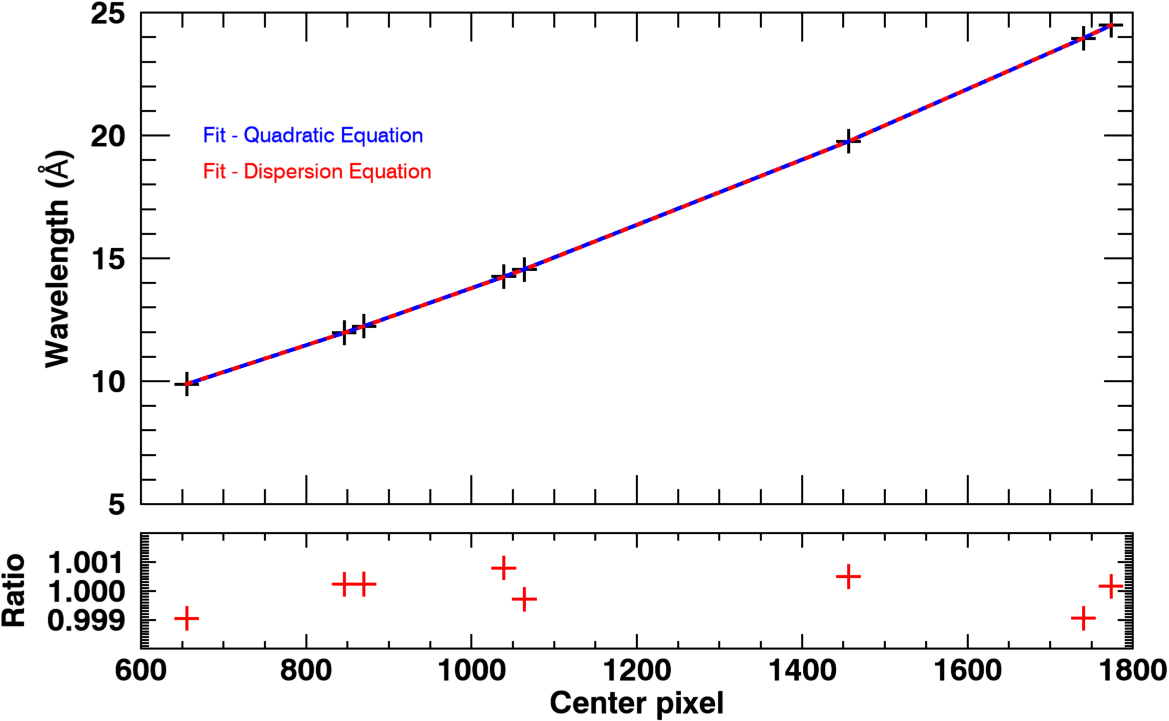}
    \includegraphics[width=0.5\textwidth]{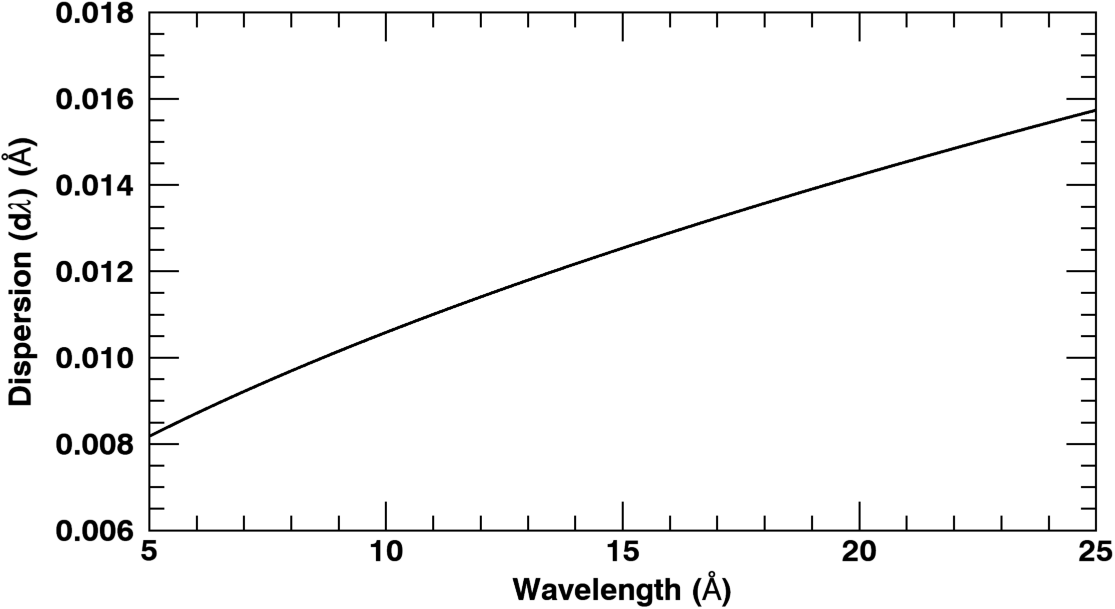}
   \caption{(Left) Wavelength calibration of {\magixs} using eight X-ray lines from three targets including first and second order spectra modeled using a quadratic equation and grating dispersion equation.  The bottom panel shows the measure of goodness of the model fit by taking the ratio of calibration wavelength to the modeled wavelength. (Right) Dispersion of {\magixs} grating showing the spectral plate scale.}
    \label{fig:calib_dispersion}
\end{figure}

 The grating disperses convergent X-rays into respective first and second order spectra, which are distinctly observed at different abscissa pixel locations on the detector. The wavelength calibration defines a relationship between the wavelength of the dispersed photon and the most probable pixel location on the detector. The center pixel of the X-ray lines in the observed data was found by fitting a Gaussian curve to each spectral line. Eight wavelengths from three targets, including the first and second orders lines, listed in Table \ref{tab:calib_lines} are used for wavelength calibration. First, we have performed an empirical fit to the wavelength calibration using a second order polynomial, which is shown in Figure \ref{fig:calib_dispersion} (Left) as a blue line for the on-axis S0 position. This fit will be applied to the flight data for scientific analysis. Similar empirical fits are performed for S1 and S2 positions, and we did not find any variation from the quadratic relation for the wavelength calibration. 
 
 Using the best fit coefficients, we first converted the pixels along the dispersion direction to wavelength. We then determined the 
 spectral plate scale as a function of wavelength, described as the number of wavelengths per unit pixel, which is shown in Figure  \ref{fig:calib_dispersion} (Right). We find a spectral plate scale of 10 m{\AA} to 14 m{\AA} in the key {\magixs} wavelength range from 10 to 17 {\AA}.
 
 To understand the deviation from linearity, we model the expected center pixel location along the image plane for the wavelength of the calibration lines and fitted the curve using standard grating equation, i.e.,
 \begin{equation}
m\lambda=d(sin~\alpha - sin~\beta)
\label{eq:dispeq}
 \end{equation}
where m is the diffraction order, $\lambda$ is the  wavelength of the calibration line,   $\alpha$ is the normal incident angle,  d is the line spacing at the center of grating, and $\beta$ is the dispersion angle measured from the normal to the grating surface. We used $\alpha$ and the pixel offset as free model parameters. The best fit yields $\alpha$ = 88.3 {\degree}, which differs from the nominal optical design value of 88~{\degree}, and an offset of 76 pixels. While we recognize the deviation in the best fit $\alpha$, this could be interpreted from the instrument build up and alignment tolerances, which will be described in a future {\magixs} instrument and alignment paper (Champey et al in preparation). The resulting fit is overplotted in Figure \ref{fig:calib_dispersion} (Left) as a red line.  The goodness of the model fit is represented as the ratio between calibration wavelength and modeled wavelength, as shown in the bottom panel of Figure \ref{fig:calib_dispersion} (Left).   Hence, we confirm that the measured grating dispersion is consistent with the prescribed instrument design values.

\begin{deluxetable}{llcc}
 \tablecaption{List of target lines and respective wavelengths from the first and second order diffraction used for {\magixs} calibration.
 \label{tab:calib_lines}}
\tablehead{\colhead{S.No} & \colhead{Target Line} & \multicolumn{2}{c}{Calibration wavelengths at} \\
                          &                                  & \multicolumn{2}{c}{different diffraction orders}\\  
                          &                                  &  \colhead{I$^{st}$ Order ({\AA})}                 &  \colhead{II$^{nd}$ Order ({\AA})}}
                  \startdata
                1 &     Mg-K-$\alpha$ & 9.88                    & 19.77\\
                2 &      Zn-L-$\beta$ & 11.97                   & 23.95\\
                3 &     Zn-L-$\alpha$ & 12.25                   & 24.50\\
                4 &     Ni-L-$\beta$  & 14.26                   &       \\
                5&      Ni-L-$\alpha$ & 14.55                   &       
\enddata
\end{deluxetable}

{}

\subsection{Calibrated Spectral Image}
\begin{figure}[h]
    \includegraphics[width=0.5\textwidth]{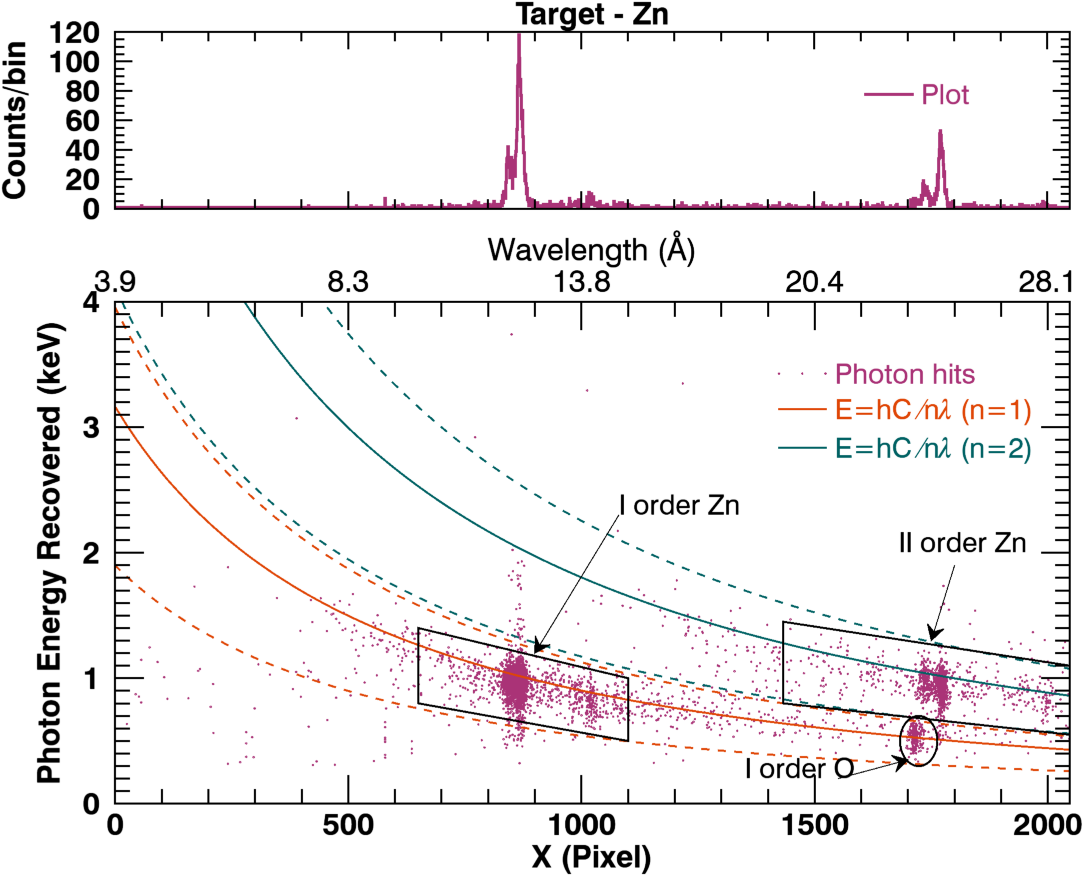}
    \includegraphics[width=0.5\textwidth]{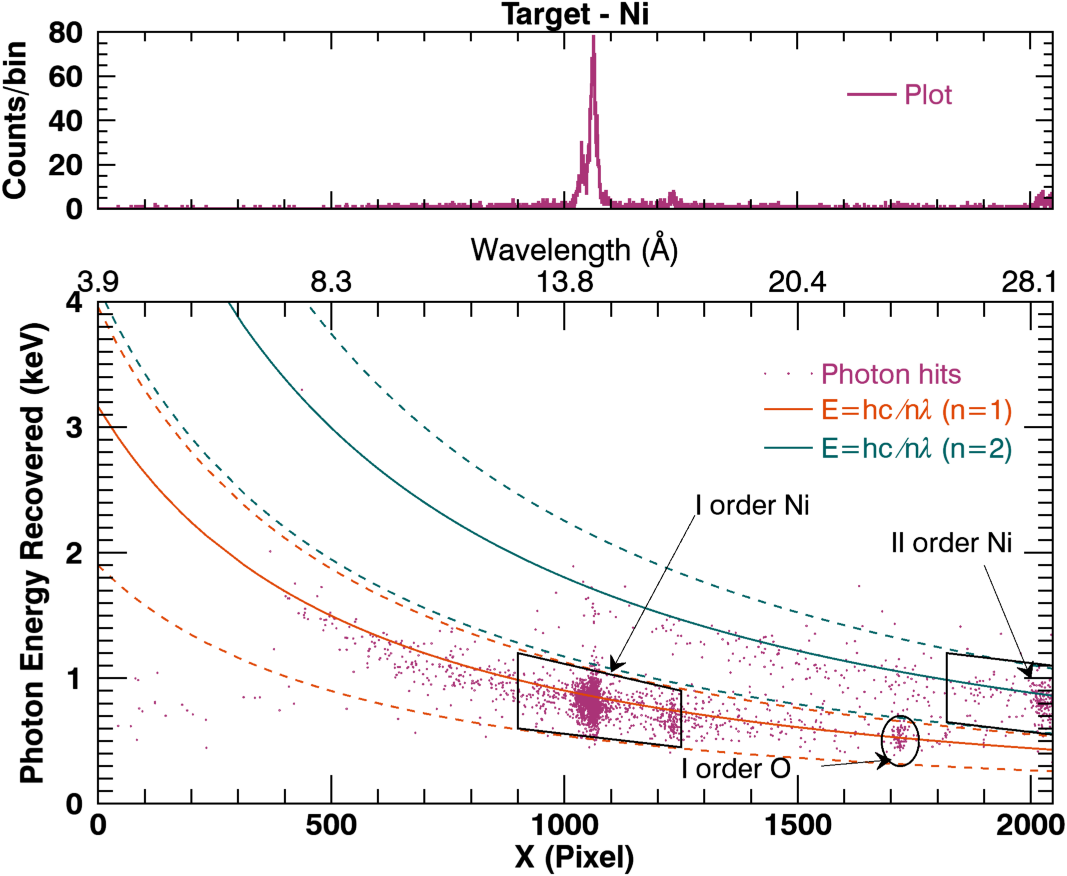}

    \caption{(Bottom Panel) Calibrated {\magixs} grating produced spectral image plot constructed for target Zn (Left) and Ni (Right) at S0 beam position on the slot.  Each dot represents a photon hit with dispersed wavelength in abscissa plotted against the respective energy deposited on the detector in ordinate.  (Top Panel) The resulting spectrum along the dispersion direction. Note that all of this spectral structure is unresolved by the BND, Figure \ref{fig:BND_marchfits}. K-$\alpha$ and K-$\beta$ lines of the respective targets are observed distinctly. The second order emission is very well observed within the detector pixels for target Zn, however it is near the edge of the detector for target Ni.}
    \label{fig:spectralimage}
\end{figure}

Now that we have an established wavelength calibration, we then create a calibrated spectral image plot with the photon energy deposited plotted against the wavelength calibrated pixels. Figure \ref{fig:spectralimage} shows the spectral image plot constructed for targets Zn and Ni at S0 beam position. Each dot in Figure \ref{fig:spectralimage} represents a photon of a particular wavelength appearing at appropriate pixel location abscissa on the detector. The energy of each photon event recovered from the algorithm is plotted in the ordinate.  The location of first and second order lines including the satellite lines are highlighted appropriately in Figure \ref{fig:spectralimage}. The solid lines are photon energy calculated from wavelength calibration for the first and second orders respectively. It is evident the target emission lines from first and second orders are well pronounced in the spectral image and fall exactly on the respective calibrated solid lines. We also notice the distinct observation of first order Oxygen peak from the target, which overlaps with the second order Zn lines, however could be well separated due to order sorting. All the other dots at high and low energies are continuum emission from the source. The target spectral lines appeared to have a spread in the recovered photon energy with an extended low energy tail. We think this could be due to the detector's inherent complex spectral response owing to charge losses and/or limitation from event processing and therefore should be included while calculating the measured flux. 

{}

\subsection{Line Spread Function}
\begin{figure}
    \centering
     \includegraphics[width=0.75\textwidth]{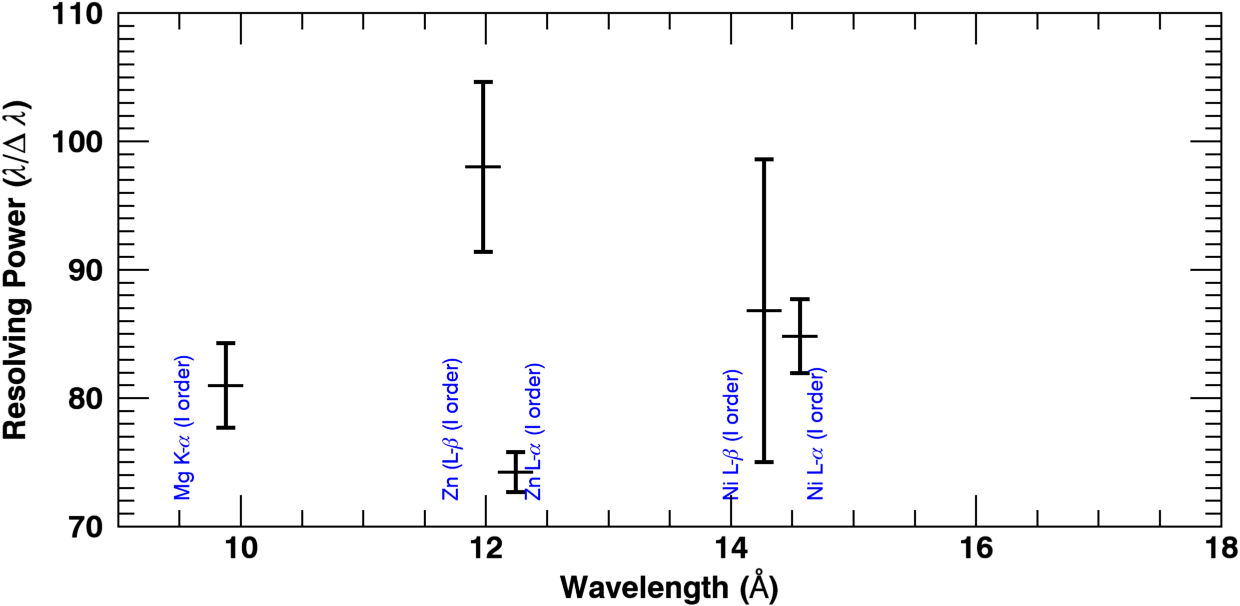}
    \caption{ The resolving power of {\magixs} determined from the line spread function (LSF) of all imaged first order lines.}
    \label{fig:linespreadfunction}
\end{figure}

As {\magixs} is an imaging spectrograph instrument, the Line Spread Function (LSF) could be envisaged as the point spread function in the spectral direction. The LSF involves convolution of the telescope and spectrometer mirrors' performance, the grating's response, and the detector's spectral response. The combined response of the X-ray mirrors strongly depend on the figure quality of the mirrors, scattering on the mirror surfaces, and co-alignment of the TMA-SOA arrangement. The flatness, groove profile, and blaze angle determine the response of the grating. The LSF determines the overall spectral resolution of the instrument, which tells how well the instrument can separate two closely spaced emission lines. We modeled the first order spectral lines with Gaussian functions and determined the Full Width at Half Maximum (FWHM - 2.35 $\times \sigma$) as a function of wavelength.  Figure  \ref{fig:linespreadfunction} shows the resolving power of {\magixs} defined by $\lambda$ / FWHM($\Delta \lambda$). The average spectral resolution of {\magixs} is $\approx$ 150 m{\AA} in the key {\magixs} wavelength range from 10 to 17 {\AA}. 

\subsection{On-axis Point Spread Function}

The point spread function (PSF) describes the instrument's ability to produce an image of a point source. For a PSF that follows the behavior of a Gaussian distribution, the width of the PSF determines the angular resolution  of the instrument. Sources closer than this value cannot be distinctly resolved. Another term for specifying the quality of the PSF is half-power diameter (HPD), or sometimes referred to as half-energy width (HEW) - the diameter at which 50$\%$ of the detected power is encircled. These terms are useful for stating the resolving limit of imaging systems that produce PSF shapes that deviate from well-behaved Gaussian functions. Here we use both Gaussian width and HPD to model and characterize the on-axis PSF produced by {\magixs}.  

\begin{figure}
    \includegraphics[width=0.5\textwidth]{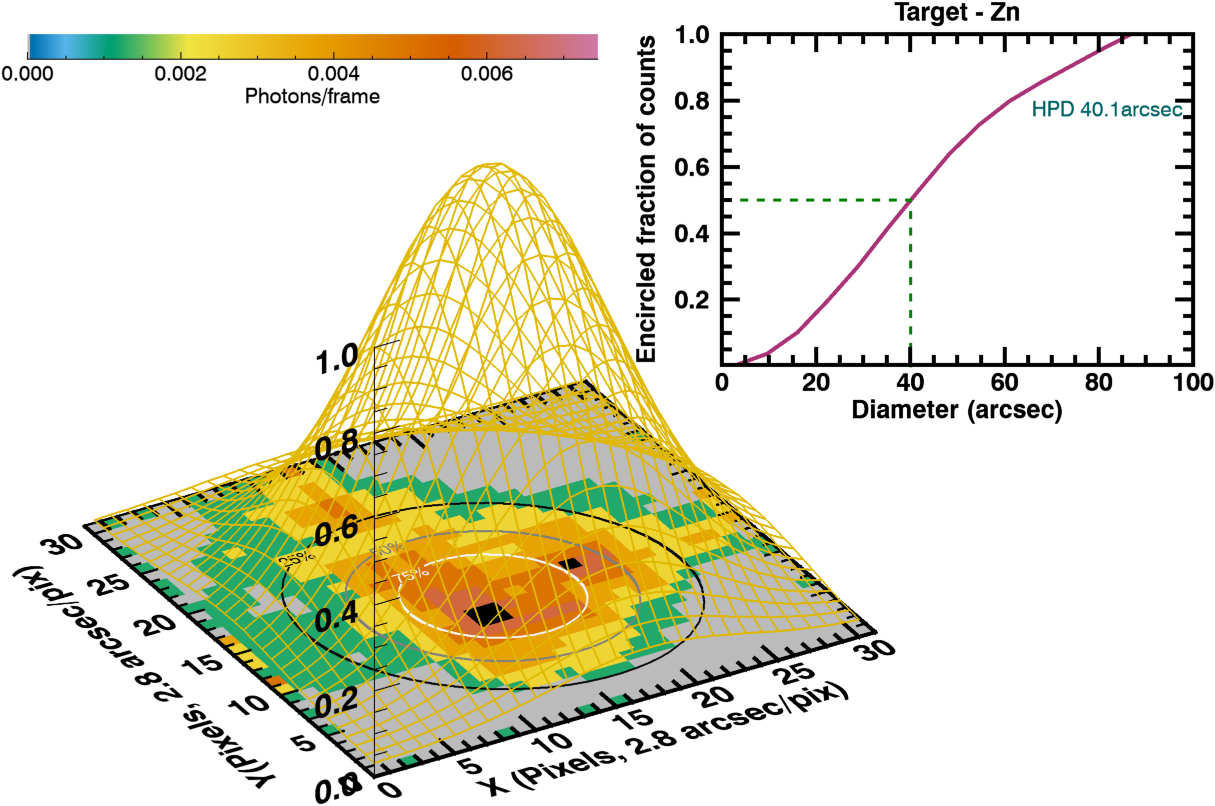}
     \includegraphics[width=0.5\textwidth]{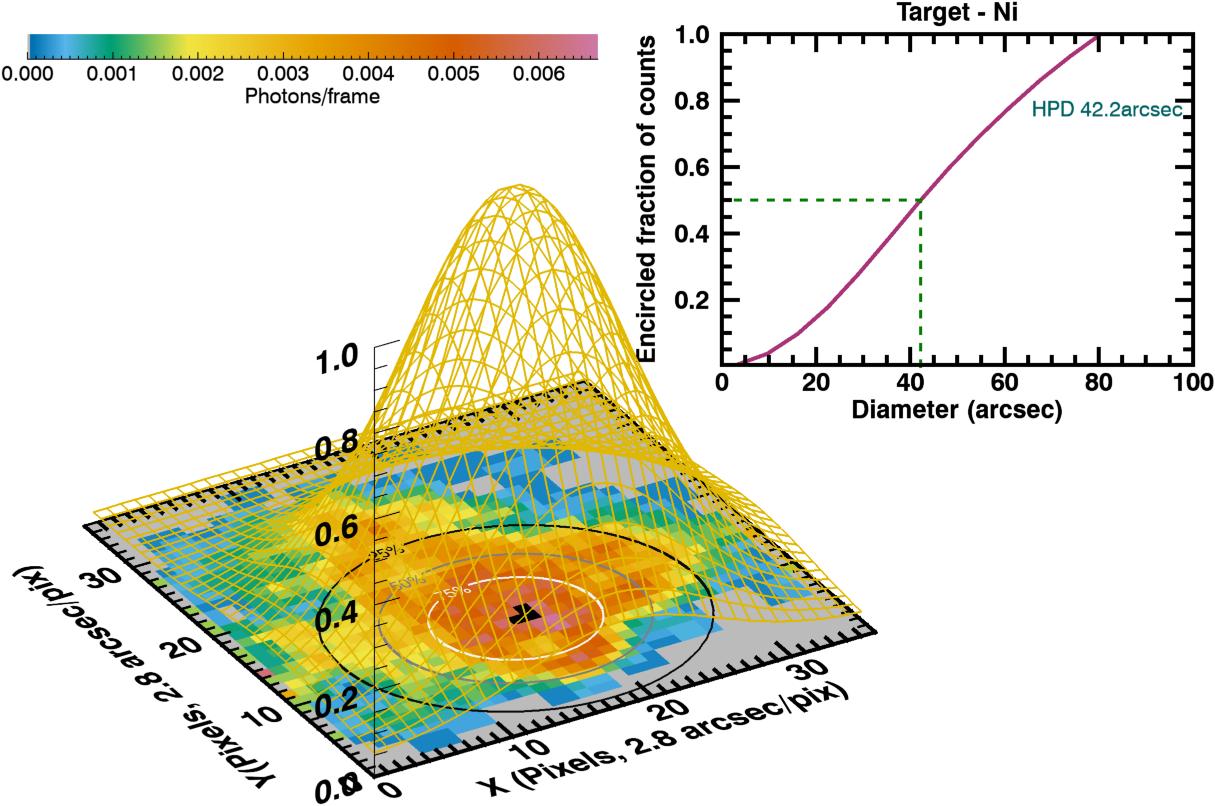}
    \caption{The measured on-axis PSF of {\magixs} for targets Zn (LefT)and Ni (Right). The PSF images are smoothed with a 3 pixel width for better view. The ellipses demonstrate the action of the modeling the PSF using a 2D Gaussian, revealing the morphology. The ellipses shown here correspond to 30\%, 50\% and 80\% of the maximum intensity in the image. The bottom left panel shows the enclosed energy fraction of the on-axis PSF. }
    \label{fig:PSF}
\end{figure}

The X-ray performance of the individual {\magixs} mirrors showed  a measurable improvement in the PSF of the active aperture (the sub-aperture dispersed by the grating), after a significant reduction of slope error was achieved using a CNC deterministic polishing technique on the mandrels \citep{champey2019, Davis2019}. \citet{champey2019} estimated the combined on-axis, full-aperture PSF using the on-axis image data from each of the three individually tested mirrors comprising the TMA and SOA. However, the combined PSF of the small 34{\degree} aperture dispersed by the grating was omitted. Using the calibration data, we directly measured the combined on-axis PSF of {\magixs}. The resulting PSF includes the response of telescope mirror, spectrograph mirrors, the grating performance at different wavelengths, and both residual and gravity induced misalignment. We modeled the PSF using a 2-dimensional Gaussian function as shown in Figure \ref{fig:PSF}. The best fit FWHM in the cross-dispersion direction is about 42 - 47 \arcsec{}. Using the centroid determined from the best fit, we measured the  HPD. The bottom  left panel in Figure \ref{fig:PSF} shows the plot of encircled normalized  counts as a function of circle diameter. In order to determine the cutoff radius for HPD calculation we considered a conservative radius value at which the gradient in the encircled flux is less than 10\%. From Figure \ref{fig:PSF} we measured the HPD of {\magixs}, the purple dashed line, which is about 40 - 45 \arcsec{}.  

For cases when the PSF has a sharp core, and narrow scattering wings, the curve tends to flatten quickly as the integral approaches the total integrated counts. However, our calibration experiments show that the PSF does not have a well defined core, and the scattering wings are broad and diffuse. For this reason, the integrated count rate does not flatten quickly over the size of the evaluated frame. We hypothesize that this can be explained by the effects of the optics assemblies deflecting with respect to each other, under a 1G load. During the experiment, the instrument roll angle was positioned such that the cross-dispersion plane (Y axis) was oriented parallel with the gravity vector. This would imply the influence of 1G deflection on the PSF would dominate in the cross-dispersion (Y) axis, which is consistent with the observed PSF shape. Dynamic and static analysis of the optical bench mechanical design indicated that a 1G environment causes a small, but significant amount of deflection between the TMA and SOA. We are continuing to investigate the source of this elongation of the spot in the Y (cross-dispersion) axis, and are applying the knowledge gained from our alignment measurements and mechanical analyses to the ray trace model, so that we can evaluate the influence of gravity induced misalignment on the PSF. 

\subsection{Radiometric Calibration}

\begin{figure}[h]
    \includegraphics[width=0.5\textwidth]{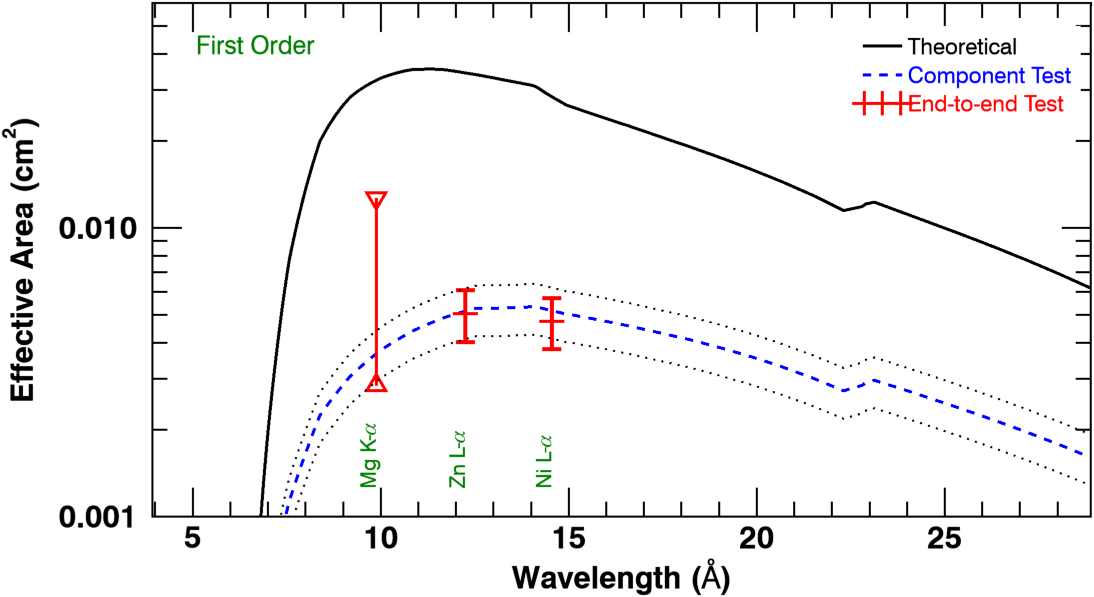}
    \includegraphics[width=0.5\textwidth]{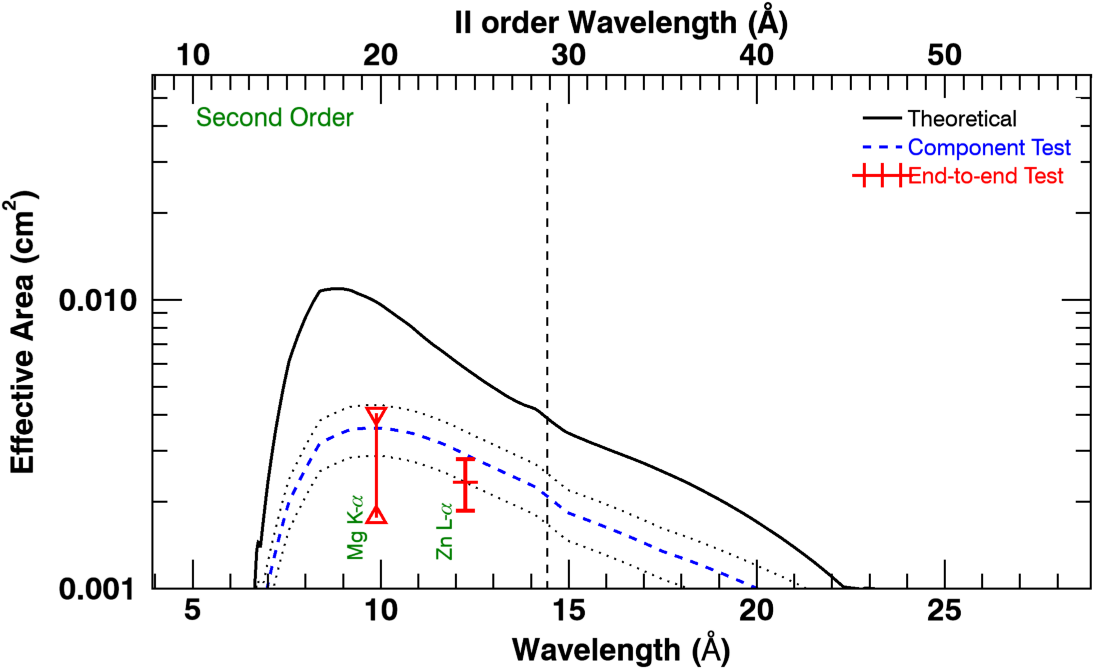}
    \caption{The effective area of {\magixs} instrument for the first (Left) and second order (Right) respectively. Only a portion of the second order spectrum falls on the detector. The vertical dashed line  on the right panel denote the wavelength corresponding to the last pixel on the detector. The solid  black lines show the effective area calculated analytically using standard X-ray database values, excluding the entrance filter. The dashed blue lines represent the effective area updated with radiometric corrections calculated from component level testing, derived from the combined reduced reflectivity of TMA, SM1, SM2, and measured grating efficiency.  The first order effective area derived from the component level tests is $\sim$ 4.5 times less than the theoretical values.  The red points are the measured effective area at the characteristic X-ray wavelengths of the calibration targets Mg (9.88 {\AA}), Zn (12.25 {\AA}) and Ni (14.25 {\AA}). The error bars are derived from the propagation of photon noise and the level of uncertainty in the incident X-ray flux. Due to additional noise in the Mg data, only a meaningful upper and lower bounds are possible to determine.  There is good agreement between the effective area derived from component level measurements and effective area from calibration tests.}
    \label{fig:ea}
\end{figure}

The aim of radiometric calibration for {\magixs} is to estimate the actual effective area of the end-to-end system.  We can do this several ways: 1) theoretical calculations, 2) component level tests, and 3) end-to-end tests of the full system.  In this section we present the  results from these three methods.  The effective area is a wavelength dependent quantity.  Component and calibration tests provide measurements at discreet wavelengths only.  We combine the theoretical and tests values to predict the effective area at all {\magixs} wavelengths. The results, summarized in Figure~\ref{fig:ea} show that both the end-to-end tests and the component level tests indicate the effective area of {\magixs} is reduced to roughly 20\% of the theoretical value.

\subsubsection{Theoretical Effective Area}

The theoretical effective area ($A_{eff}$) of {\magixs} in flight configuration is defined by the following equation
\begin{equation}
        A_{eff} =  T_{Ent} \times 
        (A_{geo} \times
        R_{primary} \times
        R_{secondary} )\times
        (R_{SM1} \times 
        R_{SM2} \times 
        G_{eff})\times 
        T_{FP} \times 
        QE_{CCD}
        \label{eq:eaeq}
\end{equation}
where $A_{geo}$ is the geometric effective aperture area in cm$^{2}$, $T_{Ent}$ is the transmission of the entrance filter, $R_{primary}, R_{secondary}, R_{SM1}$, and  $R_{SM2}$ are the reflectivities of two surfaces of the Wolter-I, SM1 and SM2 optics, $G_{eff}$ is the measured first order efficiency of grating (see sec 4.6.1), $T_{FP}$ is the transmission of focal-plane filter, and $QE_{CCD}$ is the quantum efficiency of the CCD camera. All values in Equation~\ref{eq:eaeq} except the geometric area are unitless and reflect the percentage transmission (in case of filters), reflection (in case of optical elements), or detection (in case of CCD).  Except for the geometric area, all terms are functions of wavelength. In test configuration, the entrance filter was not used, and the theoretical effective area calculated is shown in Figure~\ref{fig:ea} (solid black line). For the quantum efficiency of CCD ($QE_{CCD}$), we assumed the values of {\xrt} CCD. The reflectivity of X-ray mirrors including TMA, SM1, and SM2, and the X-ray transmission through window materials were calculated using the optical prescription values given in Table \ref{tab:opticalcomponents}, following the tabulated X-ray database values from Center for X-ray Optics (CXRO) {\bf https://henke.lbl.gov/optical\_constants/}.  The theoretical efficiency of grating was calculated using PCGrate \citep{goray_2006} and the parameters listed in Table~\ref{tab:opticalcomponents}, is shown as solid lines in Figure \ref{fig:gratingeff}. The first and second order theoretical effective area calculated using the values described here is given in Figure~\ref{fig:ea} as black lines. 

\subsubsection{Component Level Testing\label{sec:compcal}}

In this section, we discuss the component level testing that was completed on optical components before integration (the grating) or during optical alignment (the TMA and SOA without the grating).  There were no component level measurements of the focal plane filter or CCD, so we include their theoretical values when calculating the effective area.  

\begin{figure}
    \centering
    \includegraphics[width=0.5\textwidth]{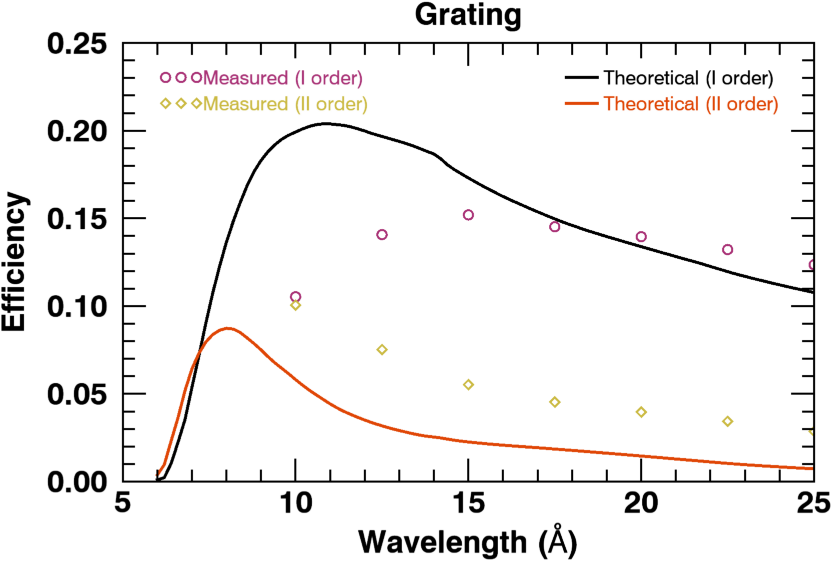}
    \caption{The first and second order theoretical efficiency of the grating measured are shown with solid black and red lines.  The measured values of efficiency for the first order at different wavelengths are shown in circles, and the second order efficiency shown in diamonds.}
    \label{fig:gratingeff}
\end{figure}

Before integration into the SOA, the first and second order grating efficiency were measured using the Advanced Light Source beamline 6.3.2 at Lawrence Berkeley National Laboratory at seven wavelengths from 10 to 25 {\AA}. The incident angle was fixed at 2 degrees and the intensity of the diffracted orders was measured by scanning a detector.  Measurements were performed at the center and +/- 25 mm from the center of the grating.  The measured values are shown along with the theoretical values in Figure~\ref{fig:gratingeff}. 

 \begin{figure}[h]
     \includegraphics[width=0.4\textwidth]{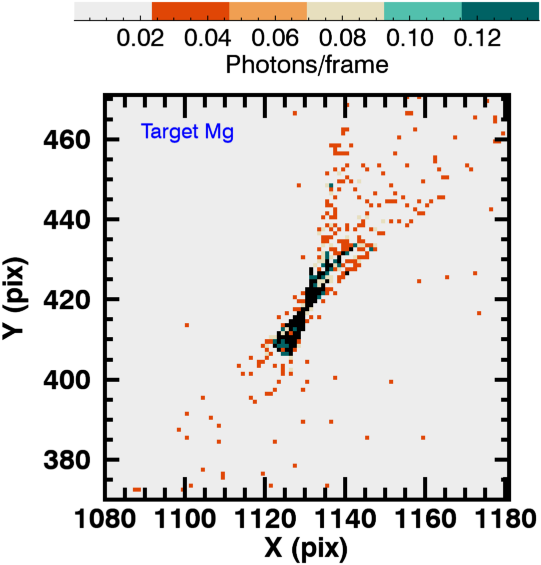}
     \includegraphics[width=0.55\textwidth]{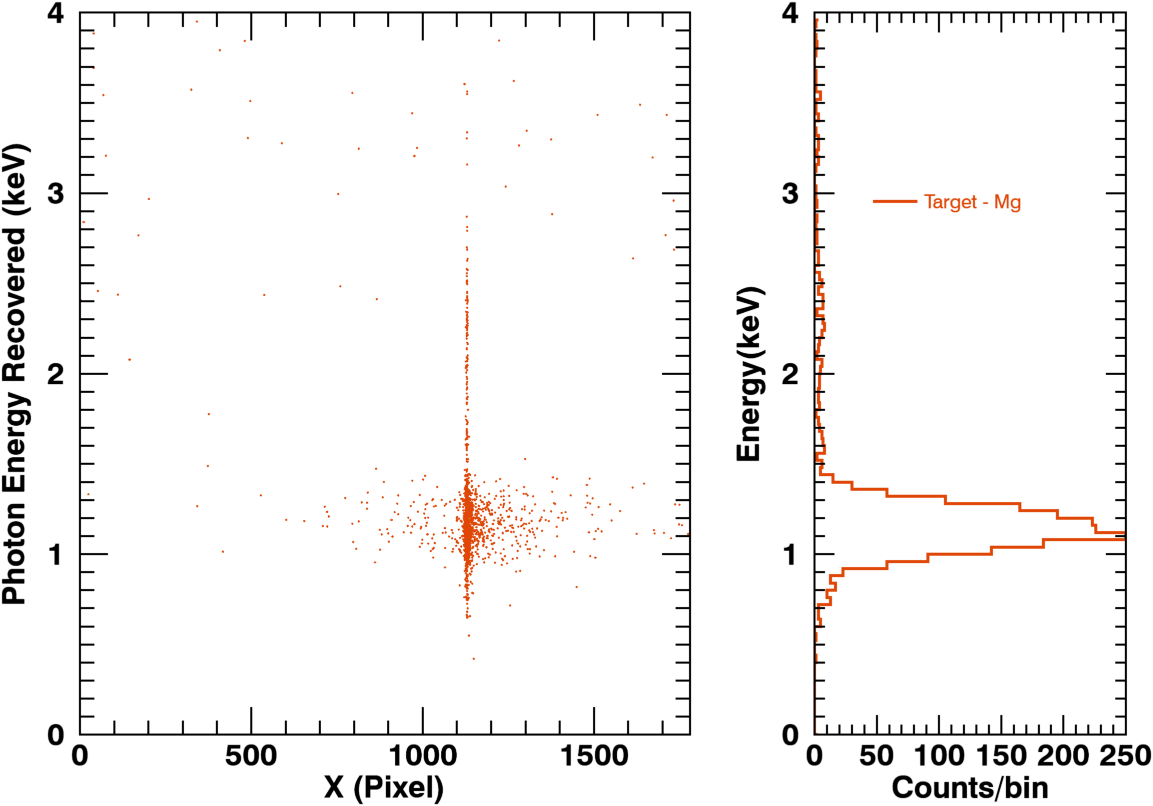}
     \caption{(Left) X-ray image of TMA sub-aperture at the best focus position. (Right) X-ray spectral image with photon energy plotted as a function of pixel location in the X-direction. The histogram of the binned energy spectrum shows the strong Mg K-$\alpha$ line in the subpanel. }
     \label{fig:tma_piece}
 \end{figure}
 
 To evaluate the performance of telescope and spectrometer mirrors we analyzed X-ray test data collected during the assembly (S.No 4 and 5 in Table \ref{tab:configs}), when the the full instrument was not yet built to compare the flux measured with the expected values. First we assessed the reflectivity of the telescope mirror from data obtained during the TMA focus test, then we determined the reflectivity of the spectrometer mirrors during the TMA to SOA X-ray alignment confirmation.   During this test, the grating was not installed. For all assembly tests, Mg target was used at the source end with respective filter with an optical depth of $\sim$ 2, which preferentially transmit the target's characteristic X-rays and suppress the continuum.  The BND data was processed as described in Paper I.
 
The presence of filter at the X-ray source significantly suppressed the continuum emission, while retaining adequate line flux. The throughput of TMA sub-aperture was measured by illuminating only the CNC polished area and the CCD kept at the best focus position, where the slot is being placed.  The CNC-polished region of the mirror could provide higher reflectivity due to reduced slope errors in the parent mandrel fabrication. The CNC-polished region of the mirror could provide higher reflectivity due to reduced slope errors in the parent mandrel fabrication.  We performed photon counting on the CCD data using the event selection algorithm (see Paper I), and measured the TMA throughput flux. Figure \ref{fig:tma_piece} (Left) shows the X-ray image of TMA sub-aperture at the best focus position; the corresponding photon counting per event and the X-ray spectrum to derive the measured X-ray flux are shown in Figure \ref{fig:tma_piece} (Right). The expected throughput flux from TMA sub-aperture is calculated by multiplying theoretical reflectivity of TMA (R$_{TMA}$) with the incident X-ray flux at TMA entrance derived from BND spectral analysis. The ratio of measured and expected TMA flux is determined to be 0.4, which is the correction factor required for the radiometric TMA reflectivity. This offset implies that the throughput of TMA reduces the intensity of the incident beam by about 60\%.

\begin{figure}[h]
     \includegraphics[width=0.45\textwidth]{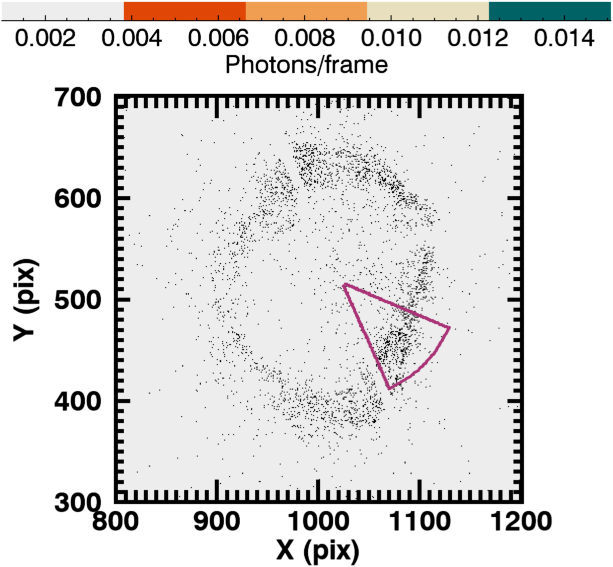}
     \includegraphics[width=0.55\textwidth]{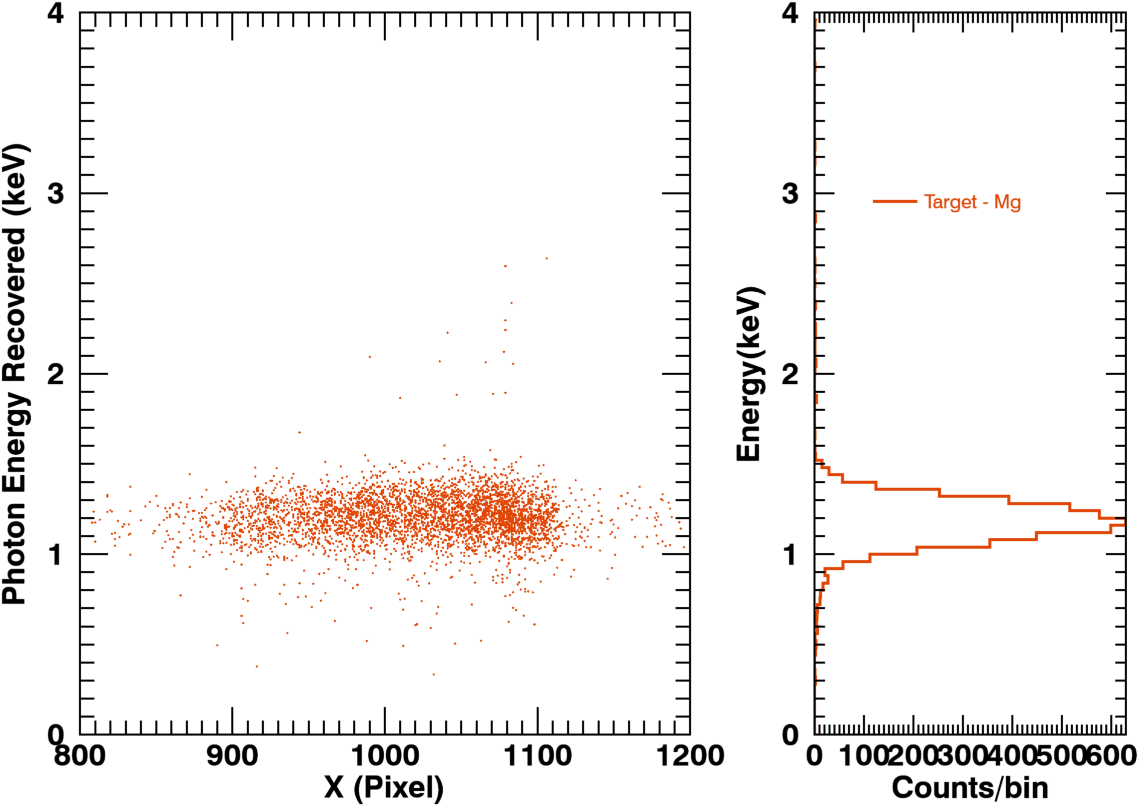}
     \caption{(Left) Full-aperture X-ray image of co-aligned TMA-SM1-SM2 configuration at an out-of-focus distance. X-ray photons coming from the CNC polished area is highlighted. (Right) X-ray spectral image of the full-aperture with photon energy plotted as a function of pixel location in the X-direction. The histogram of the binned energy spectrum shows the strong Mg K-$\alpha$ line in the subpanel.}
     \label{fig:TMA_SM1_SM2}
 \end{figure}

We then assessed the combined radiometry of spectrometer mirrors (SM1 and SM2) using assembly test data repository with co-aligned TMA, SM1 and SM2 mirror configuration at an out-of-focus position. Figure \ref{fig:TMA_SM1_SM2} shows the X-ray image of the TMA-SM1-SM2 data with the good portion of the image highlighted. The expected throughput from TMA-SM1-SM2 configuration is calculated by multiplying the incident X-ray flux at TMA entrance derived from BND spectral analysis with the theoretical reflectivity of TMA, TMA correction factor (0.4), and theoretical reflectivity of SM1, and SM2. The ratio of measured and expected flux gives the correction factor required for the combined SM1-SM2 reflectivity, which is found to be 0.54. This offset implies that the spectrometer mirrors further reduce the intensity by about 46\%. 

Therefore, the overall radiometric correction factor for the combined reflectivity of TMA, SM1 and SM2 mirrors is $\sim$ 0.216, which means that the throughput flux reaching the grating would be $\sim$20\% of the incident X-ray flux. Using the component level calibration tests we revise the theoretical effective area curve for first and second orders by including the measured grating efficiency and multiplying the optics reflectivities by the correction factors, which is shown as blue dashed line in Figure \ref{fig:ea}. Though the correction factors were taken at a single wavelength, we assumed the correction factor applies to the entire wavelength range when calculating component level effective area.

 \subsubsection{Measured Effective Area}
    \label{sec:ea}

\begin{figure}[h]
    \includegraphics[width=0.5\textwidth]{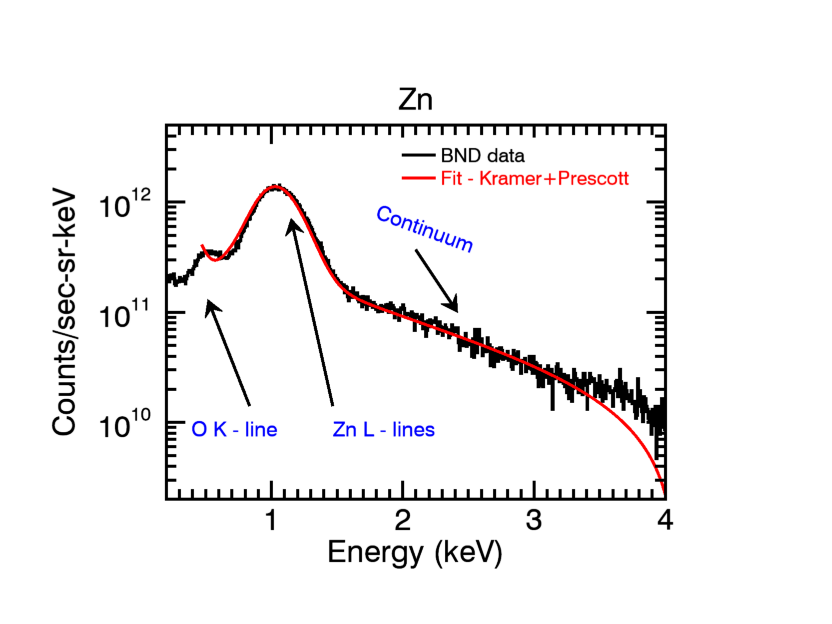}
    \includegraphics[width=0.5\textwidth]{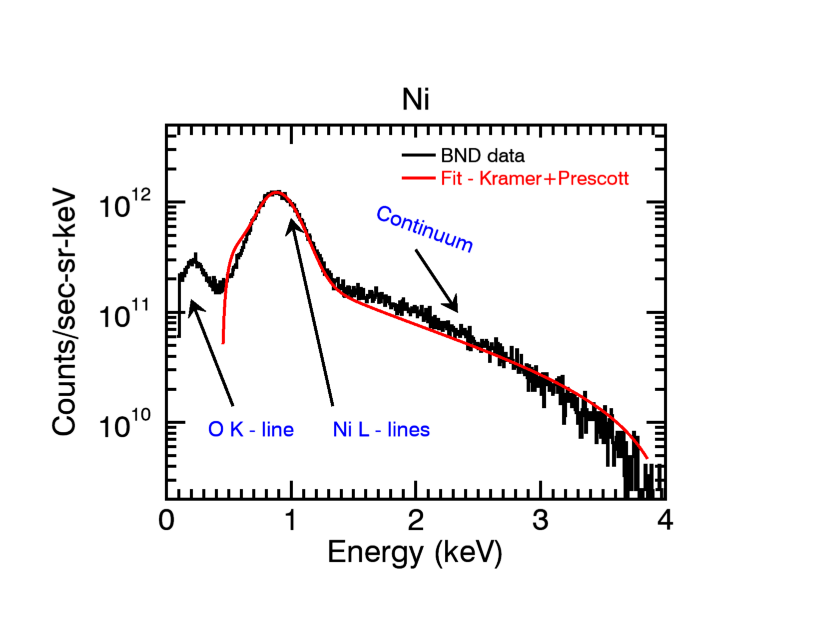}
    \caption{The unfiltered incident X-ray spectra from targets Zn and Ni as measured by the BND. The continuum and target line emission are modeled using Kramer's function and Prescott function and the corresponding line intensities are determined. We acknowledge the presence of O emission line in the observed BND spectra, however not included in spectral modeling.}
    \label{fig:BND_marchfits}
\end{figure}

We first calculate the expected incident line flux at {\magixs} aperture through spectral analysis of BND data, which is slightly different from Paper I, as there was no filter at the source end for all targets during calibration tests because we had the {\magixs} focal plane filter installed in front of CCD. Therefore, the observed BND spectra includes photons from source lines plus a strong continuum at relatively poor energy resolution.  However, {\magixs} observed a dispersed spectra with much higher spectral/energy resolution, where we can separate the continuum and the source lines.  Because the {\magixs} response varies significantly over wavelength and the goal of this exercise is to measure the effective areas at discrete wavelengths, we modeled the source line and continuum in the BND data with a Prescott function \citep{Prescott1963} and Kramer's equation. We mention that the  flux derived under the line would also have some continuum contribution ($<$10\%). Apart from the source line, we also observed the low-energy oxygen line in the BND data, however, we did not include spectral modeling to avoid complex spectral response deconvolution of BND, which is beyond the scope of this work. For spectral analysis and flux estimates, we used the quantum efficiency of BND \citep{Martin1997}. Sample spectral fits to BND data with models for line and continuum emission are shown in Figure \ref{fig:BND_marchfits}.

{}

The total number of photons reaching the {\magixs} science camera after dispersion from the grating is determined from calibrated spectral image shown in Figure \ref{fig:spectralimage}. We count the total number of photons within a tolerance energy level considering a range of pixels encompassing all the target lines, including satellite lines.
For instance, with the Zn target, we count all detected photons that fall between pixel $\sim$ 600 and 1100 and have a tolerance energy range in E$_{ph}$-35\% to E$_{ph}$+25\% E$_{ph}$ as a detected Zn photon, where E$_{ph}$ is the photon energy. The tolerance levels are marked as dashed lines in Figure \ref{fig:spectralimage}. The reason for the larger tolerance (-35\%) on the lower wavelength side is to accommodate the low energy tail photons. In addition, we also observe a small fraction of pile up photons with energy deposits greater than target line energies, appearing at the respective target's first order pixel wavelengths.  We include these photons in our flux estimates by counting them twice. We divide the number of detected photons by the total observation time to get the photons/s in the {\magixs} science camera.  We then calculate the effective area of {\magixs} at the target's line energy by taking the ratio of this to the expected flux determined from the BND analysis. This method works well for the data from the Zn and Ni targets.  

As mentioned in Section \ref{sec:setup}, the Mg data suffered from additional noise due to the cooling system pausing and the detector warming during data collection.  The noise floor of the data increased over time and ``hot pixels'' increased intensity over time. These hot pixels masquerade as photon hits and can significantly impact flux estimates. We identified columns of hot pixels and determined meaningful upper and lower flux limits by including and excluding them in photon counting, which are shown as triangles in Figure~\ref{fig:ea}.

\subsubsection{Results of Radiometric Calibration}
Figure \ref{fig:ea} shows the effective area curve of {\magixs} for the first order (Left) and second order (Right). We mention that only a portion of the second order spectrum falls on the detector, which would result in a systematic underestimation of flux from the measurement data. The solid black lines denote effective area calculated analytically using standard X-ray database values excluding the entrance filter using the optical prescription values given in Table \ref{tab:opticalcomponents}. The estimated effective area derived from the component level tests is shown by the dashed blue lines. This value includes measured grating efficiency (Figure \ref{fig:gratingeff}) and the correction factors derived for the reflectivity of the TMA, SM1 and SM2 mirrors measured at a single wavelength, which was then applied for the entire waveband. The measured effective area at discrete wavelengths from the calibration tests are overplotted with error bars. The error bars are calculated by propagating errors in the measured {\magixs} flux, which is statistical Poisson noise and a conservative 20\% error for the expected incident flux determined from Paper I, which dictates the final error in the measured effective area. Due to the additional noise in Mg data (see Section \ref{sec:setup}), we have derived meaningful upper and lower limits for the effective area, denoted by the triangles. We also applied a conservative error estimate of 20\% to the effective area derived from the component level tests, showing the upper and lower bounds for the deduced effective area curve, represented by dotted lines in Figure \ref{fig:ea}. The measured effective area is $\sim$ 4.5 times less than the theoretical radiometric prediction.  The effective area curve derived from the component level tests using correction factors for the reflectivity of mirrors and the measured grating efficiency closely agree with the measured effective area values at discrete wavelengths.

\section{Prediction of count rates for MaGIXS flight}
\label{sec:flightpredict}
\begin{figure}[h]
    \includegraphics[width=0.5\textwidth]{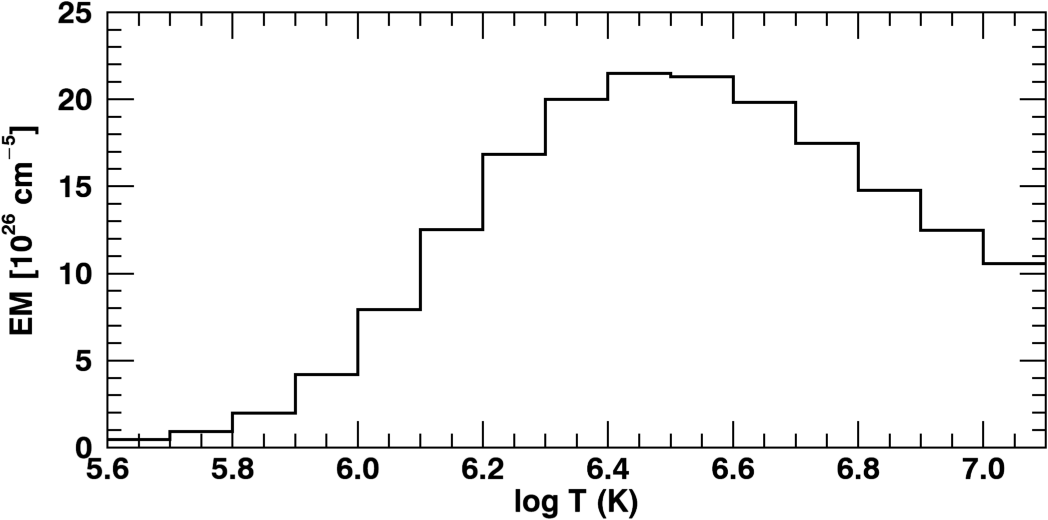}
    \includegraphics[width=0.5\textwidth]{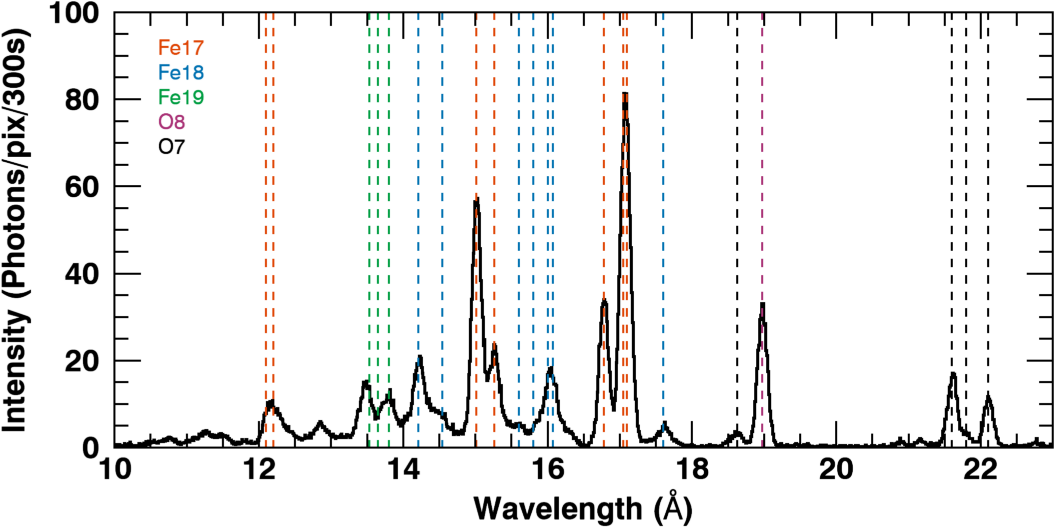}
    \caption{(Left) The emission measure distribution of a typical solar active region taken from Chianti atomic database, used to predict in flight {\magixs} expected intensity. (Right) The predicted {\magixs} spectrum for a typical solar active region. The dashed vertical lines indicate some of the strong lines of Fe XVII, Fe XVIII, and  Fe XIX that {\magixs} will observe. This spectrum includes an instrumental broadening of FHWM = 0.15 {\AA}, which is the measured average line spread function of {\magixs}.}
    \label{fig:magixspredictedspectra}
\end{figure}

We use the measured instrument response including the wavelength calibration and effective area to calculate the expected {\magixs} intensity during the flight observing a typical solar active region. To obtain the line intensities we used emission measure (EM) distribution corresponding to a typical solar active region taken as available in CHIANTI v.9 database, as shown in Figure \ref{fig:magixspredictedspectra} (left). Figure \ref{fig:magixspredictedspectra} (Right) shows an example of the predicted {\magixs} spectrum, which is obtained by folding {\magixs} response through the EM distribution shown in Figure \ref{fig:magixspredictedspectra} (Left). Table \ref{tab:predictedline} gives some of the strong emission lines in the {\magixs} wavelength range for different ion species, the maximum temperature of emission, and the expected signal strength in the units of photons/pix/300s integrated over spectral lines from first and second orders, for the entire flight duration.  We note that these EM estimates are approximate and are derived based on available space instrumentation, which has a `blind-spot' for high temperature EM slope as demonstrated in \citep{winebarger2012, Athiray2019}.  We notice the strongest and brightest lines are Fe XVII 17.07 {\AA}, 16.78 {\AA}, 15.01 {\AA}, O VII 21.60 {\AA}, 22.10 {\AA}, O VIII 18.97 {\AA}, but there are lines from Fe XVIII 14.54 {\AA}, 14.21 {\AA}, 15.83 {\AA}, 16.00 {\AA}, 16.07 {\AA} and Fe XIX 13.53 {\AA} in the {\magixs} wavelength range. These lines provide the expected temperature sensitivity of {\magixs} covering warm and hot plasma emission. We notice that for the simulated typical active region emission measure distribution, the predicted count rates for the diagnostic Fe XVIII and Fe XIX are low and therefore several pixels may need to be summed to achieve adequate counts in those emission lines.

\begin{deluxetable}{c c c c c}
\tablecaption{Predicted {\magixs} flux for some of the strong emission lines in the MaGIXS Wavelength Range \label{tab:predictedline}}
\tablehead{
\colhead{Ion} & \colhead{Wavelength}  & \colhead{Log Maximum }   & \colhead{Predicted {\magixs} flux} & \colhead{Predicted {\magixs} flux} \\
\colhead{} & \colhead{}  & \colhead{Temperature}   & \colhead{I order } & \colhead{II order} \\
\colhead{} & \colhead{{\AA}}  & \colhead{}   & \colhead{Photons/pix/300sec} &\colhead{Photons/pix/300sec}}
\startdata
O  \sc{VII} & 22.10\,\AA &6.30&  401&\\
O  \sc{VII} & 21.60\,\AA &6.30& 661&\\
O  \sc{VIII} & 18.97\,\AA & 6.40 & 847& \\
Fe \sc{XVIII}& 17.62\,\AA & 6.80 & 18 &\\
Fe \sc{XVII}& 17.07\,\AA & 6.60 & 677&\\
Fe \sc{XVII}& 16.78\,\AA & 6.60 & 275&\\
Fe \sc{XVIII}& 16.07\,\AA & 6.80 & 37 &\\
Fe \sc{XVIII}& 16.00\,\AA & 6.80 & 24 &\\
Fe \sc{XVIII}& 15.83\,\AA & 6.80 & 14 &\\
Fe \sc{XVII}& 15.26\,\AA & 6.60 & 125&\\
Fe \sc{XVII}& 15.01\,\AA & 6.60 & 432&\\
Fe \sc{XVIII}& 14.54\,\AA & 6.80 & 14& \\
Fe \sc{XVIII}& 14.21\,\AA & 6.80 & 45 &\\
Fe  \sc{XIX} & 13.53\,\AA & 6.95 & 23 &\\
Fe  \sc{XVII} & 12.2\,\AA & 6.95 & 80 & 52\\
Fe  \sc{XVII} & 11.2\,\AA & 6.80 & 15 & 11\\
Mg \sc{XI} & 9.3\,\AA & 6.80  &10 & 11\\
Mg \sc{XI} & 9.2\,\AA & 6.80  &15 & 18\\
%Ne  \sc{X} & 12.13\,\AA & 6.70 & 18 \\
%Fe \sc{XX} & 12.83\,\AA & 7.05 \\
\enddata
\end{deluxetable}
{}

\begin{deluxetable}{c c}
\tablecaption{{\magixs} Calibration Parameters. 
\label{tab:calib_parameters}}
\tablehead{
\colhead{Parameter} & \colhead{Value}}
\startdata
Pixel-to-Wavelength & 2.231 $\times$ 10$^{-6}$ (II order term)\\
calibration coefficients&0.00763 (I order term)\\
&3.93(offset)\\
\hline
Average spectral plate scale (10 - 17 {\AA}) & 11~m{\AA}\\
\hline
Average line spread function & $\sim$ 150 m{\AA} in 10 - 17 {\AA}\\
\hline
 Effective area (cm$^2$) & 0.0050 at 12.25 {\AA}\\ 
 &  0.0047 at 14.55 {\AA}\\
%  & \\
\hline
Point Spread Function & 40 - 45 arcsec HPD
\enddata
\end{deluxetable}

\section{Summary and Discussion}
\label{sec:summary}
In this paper, we have presented our analysis and results of {\magixs} end-to-end instrument calibration using the 500~m X-ray beam at XRCF at NASA MSFC. The measurements are carried out using X-ray beams from three different targets viz Mg, Zn, and Ni. Calibration data are analyzed using photon counting method published in Paper I, to measure  photon energy and flux. We have performed wavelength calibration using eight wavelengths arising from the first and second order grating diffraction of the three selected target emission lines. We find that pixel to wavelength calibration can be well modeled using a quadratic function, as shown in Figure \ref{fig:calib_dispersion}.  The measured dispersion in the key {\magixs} wavelength range i.e. from 10 to 17 {\AA} is found to be 10~m{\AA} to 15~m{\AA}. Using photon counting we then created calibrated spectral images  with the wavelength calibrated pixels plotted against the photon energy deposited on the pixels, as shown in Figure \ref{fig:spectralimage}. Using these plots we highlighted the significance of photon counting to identify emission at different wavelengths including the capability to delineate the overlap of different first and second order emission lines. We observed a close match between the photon energy derived from the wavelength calibration and the photon energy deposited on the detector. Furthermore, we performed spectral line fits to the calibration data along the dispersion direction and determined the line spread function to be $\approx$ 150~m{\AA}.

Using the simultaneous measurement of incident X-ray spectra with a BND, we measured the effective area of the {\magixs} instrument, which is  $\approx$ 4.5 $\times$ less than the theoretical effective area calculated from standard X-ray database values for the  mirrors, filter transmission, and grating efficiency.  In addition, we also analyzed X-ray data from the assembly tests with aligned TMA and TMA-SM1-SM2 configuration and evaluated the radiometric throughput of telescope and spectrometer mirrors. We find a correction factor of 0.4 to the TMA reflectivity and 0.54 for the combined reflectivity of SM1 and SM2. The combined radiometric correction factor for the throughput from the telescope and spectrometer mirrors is 0.216, which closely agrees with the measured {\magixs} effective area from the calibration tests. This offset implies  a reduction of about 80\% in effective area from the theoretical values, most of that reduction coming from the telescope and spectrometer mirrors. From our understanding of the instrument and experiment setup there could be many reasons why we see an overall reduced throughput from these mirrors. However, it is not trivial to quantify these factors. Here, we list some of the major factors of which  some combination could be responsible for the overall reduction in the throughput.
\begin{itemize}
    \item {{\bf Surface roughness of telescope and spectrometer mirrors}: {\magixs} mirrors are produced using electroforming replication process, pioneered at MSFC — the same process used in HEROES (High-Energy Replicated Optics), ART-XC (Astronomical Roentgen Telescope – X-ray Concentrator), FOXSI (Focusing Optics X-ray Imager) and IXPE (Imaging X-Ray Polarimetry Explorer). The prescribed surface roughness of the X-ray mirrors are $\leq$ 5 {\AA}. At this point, the mandrels used for replicating {\magixs} mirrors are fabricated using the state-of-the-art CNC polishing technique. The surface profile of the CNC polished mandrels mapped using metrology showed a significant reduction of slope errors and the estimated half power diameter ranges $\approx$ from 3.5" to 7". However, an important caveat is that there is no direct way to measure the surface profile of the replicated mirror shells. The X-ray performance evaluation of these mirrors showed only a marginal improvement in the HPD. Historically, the replication errors are expected to degrade the performance by a factor of two, which sets an approximate upper limit of 10{\AA} to the surface roughness. The theoretical reflectivity of the TMA Wolter-I mirror calculated using standard X-ray database values for the optical prescription given in Table \ref{tab:opticalcomponents} is shown in Figure \ref{fig:TMAreflectivity} represented by the solid line. Keeping the optical prescription the same, we vary the surface roughness of the mirror until the TMA reflectivity is $\approx$ 40\% of the original values, plotted as the sequence of dotted lines in Figure \ref{fig:TMAreflectivity}. The plot indicates that a surface roughness of $\approx$ 4~nm is required if that is to solely account for the reduced reflectivity. This is an extreme case and is not supported by the measured on-axis optical performances. Therefore, the real surface roughness of the mirrors, which we cannot directly measure at this point, likely lies somewhere between these two bounds.}
    \begin{figure}[h]
        \centering
        \includegraphics[width=0.5\textwidth]{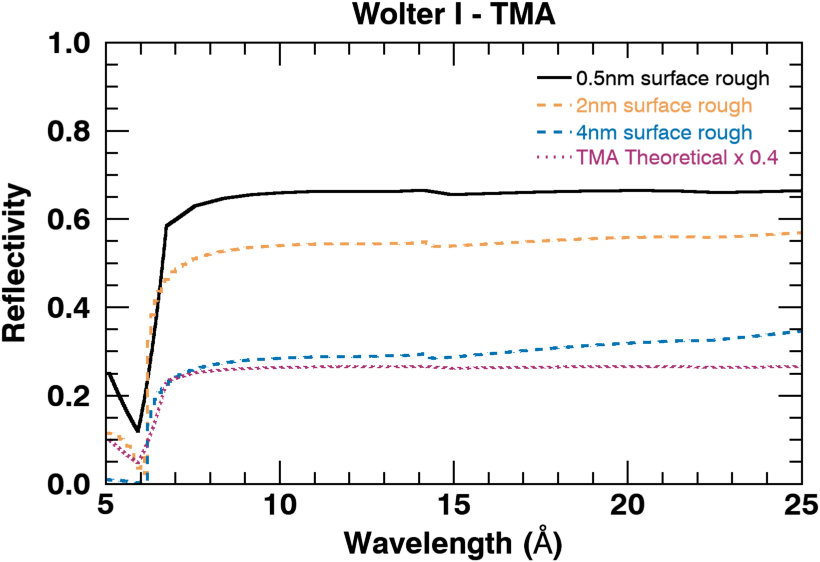}
        \caption{Reflectivity of TMA Wolter-I type mirror, including primary and secondary for different surface roughness values. Solid curve represents reflectivity for the mirror prescription given in Table \ref{tab:opticalcomponents}. Surface roughness of $\approx$ 4~nm (dashed line), an extreme case, is required if it were to explain the reduced TMA reflectivity (dotted line).}
        \label{fig:TMAreflectivity}
    \end{figure}
    \item {{\bf Vignetting and Alignment errors}: In general, X-ray mirrors in a telescope are optimized around the optical axis, where the effective area is maximum. An increase in off-axis angle of the mirrors would worsen the PSF, and also affect the throughput flux. This situation is generally known as vignetting, which is defined as the ratio between the measured source intensity imaged at a given position and the measured intensity of the same source aligned to the optical axis. For {\magixs}, where the X-ray mirrors are arranged in series followed by the grating, the relative co-alignment of TMA, spectrometer mirrors, and grating with respect to optical axis is critical. The assemblies (TMA, SOA) are co-aligned on the optical bench using a theodolite up to the mechanical tolerance levels. However, when there is an offset of the optical center from the geometrical center of any of the mirror, it would result in vignetting, which is very challenging and complicated to experimentally measure and quantify it. We have performed experiments to assess if there is any   possible vignetting by the top-hat structure that supports the entrance filter of the TMA, through X-ray measurements with and without top-hat, which did not reveal any significant change in the TMA throughput. However, these results are difficult to interpret as they were conducted on different dates, and in between some changes are made to the setup.  The contribution from vignetting is not fully addressed in this first {\magixs} rocket flight calibration, and a more through investigation will be carried out for future flights. We emphasize that this will not significantly affect {\magixs} science observations as we have measured the effective area.}
    \item{{\bf Grating efficiency}: The measured first order grating  efficiency is less than the theoretical prediction for low wavelengths below $\sim$ 15~{\AA} (see Figure \ref{fig:gratingeff}) while the second order over performs by $\sim$ 100\% at all measured wavelengths. We also notice a difference in the shape of the efficiency curve between the theory and measurement. These observations could be explained by an imperfect blaze angle in the grating. For instance, an increase in blaze angle from the nominal value would shift the peak of the efficiency curve to longer wavelengths in both first and second orders respectively. This change in blaze angle would also reduce the first order efficiency considerably, while increasing the second order efficiency, as seen in the measurement. In addition, the manufacturing and coating process of the grating left residual material at the edge of each saw tooth, which could also impact the measured efficiency. A more detailed investigation of grating performance including modeling efforts will be presented in a future paper and  is beyond the scope of the work described here.}

    \item {{\bf Assumption of CCD efficiency}: We assumed the efficiency values of the {\xrt} CCD in our analysis, which is a reasonable approximation given that the {\magixs} CCD is an astro-processed, back-illuminated CCD from e2V. In \citep{Athiray2019}, we used a similar flight grade astro-processed CCD from e2V Technologies Ltd., which yields close agreement with BND results. \citep{Moody2017} showed that the astro-processed CCDs from Te2V Technologies Ltd. are demonstrated to be reliable and consistent in the {\magixs} energy range. Therefore, we conclude that although we have not directly measured the quantum efficiency of {\magixs} flight camera, we believe that it will not be significantly less as a way to explain the observed throughput of {\magixs}.}

\end{itemize}
Using the measured {\magixs} calibration products such as wavelength calibration, instrument broadening (FWHM) and the updated effective area, we predicted expected line fluxes for a typical active region observation during {\magixs} flight. However, we note that the emission measure used in our forward model is an approximation derived from the existing instruments, which cannot precisely quantify high temperature emission. Our results presented here suggest that observing bright/hot active region during {\magixs} flight would be highly beneficial to maximize science throughput.

\acknowledgements
 P. S. Athiray's research is supported by an appointment to the NASA Postdoctoral Program at the Marshall Space Flight Center, administrated by Universities Space Research Association under contract with NASA.  The {\magixs} instrument team is supported by the NASA Low Cost Access to Space program.  The authors gratefully acknowledge the many people at XRCF who have contributed to the testing. CHIANTI is a collaborative project involving NRL (USA), RAL (UK), and the following Universities: College London (UK), of Cambridge (UK), George Mason (USA), and of Florence (Italy). The authors appreciate helpful and insightful comments and suggestions from an anonymous referee. 

\bibliography{sample63,solar,references}{}
\bibliographystyle{aasjournal}

\end{document}